\documentclass[11pt,letter]{article}
\pdfoutput=1 

\usepackage{jheppub}
\usepackage{amssymb,amsmath,amsthm,graphicx}
\usepackage[english]{babel}
\usepackage{graphics, color}
\usepackage{subfigure}
\usepackage{ragged2e}
\usepackage{latexsym}
\usepackage{bm}
\usepackage[hang,flushmargin]{footmisc}
\usepackage{epsfig}
\usepackage{textcomp}
\usepackage{comment}
\usepackage{ulem}
\hypersetup{
 pdfauthor={Xiaoyi Liu, Jorge  E. Santos, Harvey S. Reall, Toby Wiseman},
 citecolor=black,linkcolor=black,urlcolor=black}
\usepackage[utf8]{inputenc}
\allowdisplaybreaks[1]

\newcommand{\be}{
\begin{eqnarray}
}

\newcommand{\nl}{
\nonumber \\
}

\newcommand{\ee}{
\end{eqnarray}
}

\newtheorem{lemma}{Lemma}
\newtheorem{prop}{Proposition}

\begin{document}
\title{\raggedright{Ill-posedness of the Cauchy problem for linearized gravity in a cavity with conformal boundary conditions}}
\author{Xiaoyi~Liu,$^a$} \emailAdd{xiaoyiliu@ucsb.edu} 
\author{Harvey~S.~Reall,$^b$} \emailAdd{hsr1000@cam.ac.uk} 
\author{Jorge~E.~Santos,$^b$} \emailAdd{jss55@cam.ac.uk}
\author{and\;Toby~Wiseman$^c$}
\emailAdd{t.wiseman@imperial.ac.uk }

\affiliation{$^a$Department of Physics, University of California, Santa Barbara, CA 93106, USA}
\affiliation{$^b$Department  of  Applied  Mathematics  and  Theoretical  Physics,  University  of  Cambridge, Wilberforce Road, Cambridge, CB3 0WA, UK}
\affiliation{$^c$Abdus Salam Centre for Theoretical Physics, Blackett Laboratory, Imperial College, London SW7 2AZ, United Kingdom}

\abstract{ 
We consider Lorentzian General Relativity in a cavity with a timelike boundary, with conformal boundary conditions and also a generalization of these boundary conditions.  
We focus on the linearized gravitational dynamics about the static empty cavity whose boundary has spherical spatial geometry. 
It has been recently shown that there exist dynamical instabilities, whose angular dependence is given in terms of spherical harmonics $Y_{\ell m}$, and whose coefficient of exponential growth in time goes as $\sim \ell^{1/3}$. 
We use these modes to construct a sequence of solutions for which the initial data converge to zero as $\ell \rightarrow \infty$ but for which the solution itself does not converge to zero. This implies a lack of continuity of solutions on initial data, which shows that the initial value problem with these boundary conditions is not well-posed. This is in tension with recent mathematical work on well-posedness for such boundary conditions.
}

\maketitle

\section{Introduction}

Boundaries have long played an important role in General Relativity (GR). York introduced boundaries in order to render the canonical partition function of black holes well defined~\cite{York:1986it}. Following from this they have played a crucial role in Euclidean semi-classical gravity with Gibbons and Hawking introducing their boundary term to study the variational problem with boundary~\cite{Gibbons:1976ue}. More recently AdS-CFT~\cite{Maldacena:1997re,Gubser:1998bc,Witten:1998qj} has introduced asymptotically Anti-de~Sitter (AdS) spacetimes with their timelike conformal boundary  playing a key role in understanding certain quantum gravities. Here York's construction is given a new perspective~\cite{Hawking:1982dh,Witten:1998zw}. In the AdS-CFT context finite boundaries play a central role. They are a regulator for the asymptotic AdS conformal boundary in holographic renormalization~\cite{Henningson:1998gx,deHaro:2000vlm,Skenderis:2002wp}, they arise in the fluid-gravity correspondence~\cite{Bredberg:2011xw,Anninos:2011zn} as well as more recently in certain CFT deformations~\cite{McGough:2016lol,Kraus:2018xrn,Hartman:2018tkw,Gorbenko:2018oov}. Numerical relativity also uses boundaries as a practical tool to regulate asymptotically flat spacetimes with absorbing boundary conditions~\cite{Sarbach:2007hd}.

While the well-posedness of the Cauchy problem in General Relativity has long been understood, the case with timelike boundaries, the `initial boundary value problem' (IBVP), remains enigmatic. Various results exist showing well-posedness for the system in a particular gauge and formulation, starting with~\cite{Friedrich:1998xt} and more recently~\cite{Kreiss:2007zz,Kreiss:2009jia,Fournodavlos:2019ckr}, but these require boundary conditions that are gauge dependent and not formulated geometrically. The task of then showing geometric uniqueness~\cite{Friedrich:2009tq} remains and this is generally an open problem for such non-geometric boundary conditions. An exception where well-posedness has been shown  for geometric boundary conditions is the case of a totally geodesic boundary~\cite{Fournodavlos:2020wde}. For an excellent review and further references on this topic see~\cite{Sarbach:2012pr}.

The Riemannian case with boundary, which is important for Euclidean semiclassical gravity, is better understood. There the obvious boundary condition is to fix the induced geometry of the boundary. Surprisingly this was shown not to give a well-posed in-filling problem~\cite{Avramidi:1997sh,Avramidi:1997hy,Anderson:2006lqb} as reviewed in~\cite{Witten:2018lgb}. Instead in~\cite{Anderson:2006lqb} Anderson proposed fixing the conformal class of the boundary together with its mean curvature (the trace of its extrinsic curvature) and showed that this gives a well-posed boundary value problem. We shall refer to Anderson's boundary condition as the conformal boundary condition. Recently a new one-parameter family of boundary conditions, which we term generalized conformal boundary conditions, was introduced in~\cite{Liu:2024ymn}. These include Anderson's conformal boundary condition as a particular case, and tend to fixing the induced metric in a limit. They are natural from the perspective of the Euclidean action as the one parameter corresponds to a rescaling of the coefficient for the Gibbons-Hawking term. These boundary conditions fix the conformal class, together with the mean curvature weighted by a power of the induced metric's determinant. 
In~\cite{Liu:2024ymn} it was shown that for a cavity filled with flat spacetime whose boundary is a product of time with a round sphere then the Euclidean Lichnerowicz operator that controls quantum fluctuations has negative modes for the conformal boundary conditions, implying that the Euclidean path integral is unstable which is physically unsatisfactory. It was also shown that there are generalized boundary conditions where it appears to be stable. The physical significance of these conformal boundary conditions for gravity is a topic of great interest, as they arise naturally in the fluid-gravity setting~\cite{Bredberg:2011xw,Anninos:2011zn}, their consequences for black holes was initially investigated in~\cite{Adam:2011dn}, and recently there are a number of works studying various aspects of their behaviour and implications~\cite{Anninos:2024wpy,Liu:2024ymn,Anninos:2024xhc,Galante:2025tnt,Banihashemi:2025qqi,Banihashemi:2024yye}. 

It is natural to wonder whether these geometric boundary conditions can be applied to a timelike boundary to give a well-posed Lorentzian IBVP. In~\cite{An:2021fcq}, An and Anderson argued that fixing the induced metric on the timelike boundary does not give a well-posed IBVP, the argument being similar to that ruling out well-posedness in the Riemannian setting~\cite{Anderson:2006lqb}.\footnote{
In very recent work it was argued that if an additional convexity condition is placed on the extrinsic curvature then the IBVP with Dirichlet boundary conditions is well-posed~\cite{An:2025gvr}. Similar arguments have been discussed in the Riemannian case~\cite{Anderson:2007jpe,Witten:2018lgb}.
}
It was conjectured there that the conformal boundary conditions do yield a well-posed IBVP. Indeed recently they have argued  that the linearized IBVP about flat spacetime is well-posed for smooth metrics with the usual Cauchy data at the initial time, together with the conformal and generalized conformal boundary conditions on a timelike boundary~\cite{An:2025rlw}. They argued that an additional ``corner angle'' between the $t = 0$ surface and the timelike boundary had to be included in this data. 

The motivation for this paper is that the claim of An and Anderson~\cite{An:2025rlw} is in tension with previous work which discovered dynamical instabilities of linearized gravity in a cavity. This was first studied in~\cite{Anninos:2023epi} for perturbations of flat spacetime with a spatial boundary that is spherical, and unstable modes were found for conformal boundary conditions. In~\cite{Liu:2024ymn} these instabilities were shown to be present also in the case of generalized conformal boundary conditions. 
This contrasts with the case of Dirichlet boundary conditions, where such cavities have been shown to be stable~\cite{Andrade:2015gja}.
It was shown that the instability is associated to the dynamics of the boundary itself because it persists even in spherical symmetry, where Birkhoff's theorem ensures that the interior geometry remains flat but nevertheless an instability still exists (even nonlinearly) and can be thought of as being due to the boundary moving in flat spacetime~\cite{Liu:2024ymn}. Furthermore, it was shown in \cite{Anninos:2023epi,Liu:2024ymn} that there exist non-spherically symmetric unstable modes, with angular dependence determined by a spherical harmonic $Y_{\ell m}$, for which at large $\ell$ the imaginary part of the frequency grows as $\ell^{1/3}$. In other words, the rate at which the instability develops increases without bound as $\ell \rightarrow \infty$.

Our work is motivated by the observation that this large $\ell$ behaviour of the unstable modes is reminiscent of what happens in the Cauchy problem for the Laplace equation, which is the standard example of a problem that is {\it not} well-posed. We will extend the analysis of \cite{Liu:2024ymn} and show that the existence of these perturbations demonstrates that {\it the IBVP for linearized gravity is not well-posed with conformal (or generalised conformal) boundary conditions}. We use a similar construction as for the Laplace equation -- namely we construct a sequence of smooth unstable modes labelled by $\ell$ such that as $\ell \to \infty$ the initial data for the sequence converges to zero but the solution itself diverges at any non-zero time. More precisely, suppose we write the unstable modes as $h_{\ell}(t)$ so that $| h_{\ell}(t) | \sim e^{\alpha \ell^{1/3} t}$ as $\ell \to \infty$ for some $\alpha > 0$. The modes are smooth, have harmonic time dependence, all obey the same boundary conditions as the unperturbed cavity, and all have the same (zero) corner angle (defined in~\cite{An:2025rlw}) as the unperturbed static cavity. Then we construct a new sequence of solutions $\tilde{h}_{\ell}(t) = e^{- \ell^{1/6}} h_{\ell}(t)$ so that initially at time $t = 0$ these limit to zero initial data, $| \tilde{h}_{\ell}(0) | \to 0 $ and $|\partial_t \tilde{h}_{\ell}(0) | \to 0 $ as $\ell \to \infty$, but for any positive time $t>0$ then $| \tilde{h}_{\ell}(t) | \to \infty$ at generic points. Hence we have a sequence of solutions for which the sequence of initial data converges to zero but the sequence of solutions does not converge to anything. This proves that the map from initial data to solutions (if it exists) is not continuous, so the problem is not well-posed. This is clearly in tension with the work of An and Anderson~\cite{An:2025rlw}. Furthermore, it means that the status of these conformal and generalized conformal boundary conditions is unclear in the Lorentzian setting, which is important when considering physical applications such as those mentioned above.

This paper is structured as follows. In Section~\ref{sec:laplace} we review an argument that shows ill-posedness of the Laplace equation for Cauchy evolution. Then in Section~\ref{sec:lingrav} we review linearized gravity in a cavity with generalized conformal boundary conditions, and the unstable modes that were previously found in~\cite{Liu:2024ymn}, rederiving them and significantly extending their previous analysis, taking particular care to confirm their smoothness. Then using the strategy from the Laplace Cauchy problem, in Section~\ref{sec:illposed} we use these modes to construct a sequence of solutions that gives an obstruction to well-posedness  for the linear gravity problem. The main technical challenge is to understand the form of these modes in the large $\ell$ limit. 
We computed the modes for the outside geometry of a spherical cavity in section~\ref{sec:outside} and those for a cylindrical cavity in section~\ref{sec:cylindrical}, finding similar instabilities to the infilling spherical cavity case. 
We conclude with a brief summary in Section~\ref{sec:summary}. There are $3$ appendices which contain details of the computations and proofs in the main text.

\section{Ill-posedness of the Laplace equation as a Cauchy problem}
\label{sec:laplace}

Consider the two dimensional Laplace equation, $\left( \partial^2_x + \partial^2_\theta \right) \phi = 0$, on the unit radius semi-infinite cylinder with coordinates $(x,\theta)$ where $\theta$ is the angle around the cylinder, so $\theta \sim \theta + 2 \pi$ and $x \ge 0$. It is well known that this problem is ill-posed as an initial value problem, treating $x$ as a ``time'' and attempting to give initial data at $x=0$:
\be
\left.\phi \right|_{x=0} = F(\theta) \; , \qquad \left.\partial_x \phi \right|_{x=0} = G(\theta)\;.
\ee
We now review Hadamard's argument that shows this, as later we will use a modification of this argument to discuss the gravitational problem. The argument is by contradiction, so assume that this initial value problem is well-posed in $C^k$ (for some $k$) or $C^\infty$. This means that there exists a unique map from smooth initial data to smooth solutions, $\xi : ( F, G ) \mapsto \phi$ and that $\xi$ is continuous in the $C^k$ (or $C^\infty$) topology on the space of initial data and the space of solutions. For some fixed $(F,G)$, with $\phi = \xi(F,G)$, consider the sequence of solutions $(\phi_n)$ defined by
\be
\phi_n(x,\theta) = \phi(x,\theta) + e^{- \sqrt{n}} e^{+ n x} \cos( n \theta ) \; .
\ee
We have $\phi_n = \xi(F_n,G_n)$ where the initial data is
\be
F_n = F +  e^{- \sqrt{n}}  \cos( n \theta ) \; , \qquad G_n = G + n  e^{- \sqrt{n}}  \cos( n \theta ) \; .
\ee
In the limit $n \to \infty$ we see that $\partial_\theta^k F_n \to \partial^k_\theta F$ and $\partial_\theta^k G_n \to \partial_\theta^k G$ uniformly in $\theta$ for any $k \ge 0$. Hence the sequence $(F_n,G_n)$ converges to $(F,G)$ in $C^k$ for any $k$ and hence also in $C^\infty$. So if $\xi$ were continuous then $\phi_n$ should converge to $\phi$ as $n \rightarrow \infty$. However the sequence $(\phi_n)$ does not have a limit in $C^k$ (or $C^\infty$) because $\phi_n(x,0)  \to \infty$ as $n \to \infty$ for any $x>0$. Thus we have proved that the map $\xi$, if it exists, cannot be continuous at the (arbitrary) point $(F,G)$ in the space of initial data. Hence the initial value problem is not well-posed in\footnote{
The same argument shows that it is not well-posed if we define continuity using a Sobolev norm instead of working in $C^k$.
} $C^k$ or $C^\infty$.

\section{Linearized gravity in a  cavity}
\label{sec:lingrav}

We now review and extend the analysis of~\cite{Liu:2024ymn}. Firstly we will introduce the setting of the gravitational cavity with timelike boundary with generalized conformal boundary conditions. We will consider an infilling of this with a perturbation of flat spacetime, where the boundary is conformal to a product of time with a round sphere.
We will review the unstable modes found in~\cite{Liu:2024ymn}, and seen previously for conformal boundary conditions in~\cite{Anninos:2023epi}. We will be careful to discuss the smoothness of these modes, and will present them in different coordinates to those in~\cite{Liu:2024ymn}.

\subsection{The setting}

We consider a ``cavity'' described by a spacetime manifold $M=\mathbb{R} \times \Sigma$ where $\Sigma$ is diffeomorphic to a closed $3$-ball in $\mathbb{R}^3$ with boundary diffeomorphic to $S^2$. The boundary of $M$ is a topological cylinder $\mathbb{R}\times S^2$. In the initial-boundary value problem we pick an initial surface in $M$, diffeomorphic to $\Sigma$, prescribe initial data on this surface and boundary data on $\mathbb{R}\times S^2$ and we wish to construct a Lorentzian metric on $M$ compatible with this data.

We will use coordinates $x^a=(t,r,\theta^I)$ where $t$ is a time function such that the initial surface is at $t=0$, $\theta^I=(\theta,\phi)$ parameterize $S^2$ in the usual way and $r$ is a ``radial'' coordinate with $0 \le r \le 1$ where $r=1$ corresponds to the boundary of the cylinder. Clearly, such coordinates are not unique. 

We will use $y^\mu=(t,\theta^I)$ as coordinates on the boundary $\mathbb{R} \times S^2$. Let $\gamma_{\mu\nu}$ be the (Lorentzian) metric induced on this boundary, and let $K_{\mu\nu}$ be its extrinsic curvature. Anderson's boundary conditions \cite{Anderson:2006lqb} correspond to prescribing $[\gamma_{\mu\nu}]$, the conformal class of $\gamma_{\mu\nu}$, along with $K \equiv \gamma^{\mu\nu} K_{\mu\nu}$. We will also consider a 1-parameter generalization of these conditions in which we prescribe the density $\gamma^p K$ on the boundary, where $\gamma=\det \gamma_{\mu\nu}$, with $p=0$ corresponding to Anderson's boundary condition.  

We will take as our background metric 4-dimensional Minkowski spacetime: 
\be
{g}^{({\rm Mink})}_{ab} = \left[ \begin{array}{ccc}
-1 & 0 & 0 \\
0& 1 & 0  \\
0& 0& r^2 \Omega_{IJ} 
\end{array} \right] \; , \qquad 
\Omega_{IJ} = \left[ \begin{array}{cc}
1 & 0 \\
0&  \sin^2{\theta}
\end{array} \right]\,,
\ee
with $\Omega_{IJ}$ the line element on the round unit 2-sphere. 
For this metric, the boundary at $r=1$ has unit radius and $K=2$. 

We will consider perturbations of this Minkowski metric. Let $\bar{\gamma}_{\mu\nu}$ denote the unperturbed boundary metric. Our boundary conditions are that $[\gamma_{\mu\nu}]=[\bar{\gamma}_{\mu\nu}]$ and
\be
\label{bcp}
\gamma^p K =  2  \bar{\gamma}^p
\ee
Note that Minkowski spacetime satisfies these boundary conditions. 
For convenience we have chosen the cavity to have unit radius. However any solution to the vacuum Einstein equation may be scaled to give another solution. That freedom here can be used to scale our perturbed cavity to have any chosen radius $R$ if we also scale the boundary condition as $\gamma^p K =  \frac{2}{R} \bar{\gamma}^p$. 

\subsection{A family of linear perturbations}

We will consider linear perturbations of the Minkowski metric. The perturbation will be assumed to have harmonic time dependence, proportional to $e^{i\omega t}$, and its angular dependence is constructed from spherical harmonics $Y_{\ell m}$ as follows. Let $\mathcal{D}_I$ be the covariant derivative on a unit round 2-sphere with metric $\Omega_{IJ}$. Spherical harmonics satisfy
\be
\Omega^{IJ} \mathcal{D}_I \mathcal{D}_J Y_{\ell m} = - \ell ( \ell + 1) Y_{\ell m} \;.
\ee
We define the symmetric traceless two tensor associated to the spherical harmonic,
\be
S_{IJ} = \left[ \mathcal{D}_I \mathcal{D}_J + \frac{1}{2} \ell ( \ell + 1) \Omega_{IJ} \right] Y_{\ell m} \; .
\ee
Now specializing to the case $m=0$ so the perturbation has no azimuthal angular dependence then we write
\be
\label{eq:sphharm}
f(\theta) = \sqrt{\frac{4\pi}{2\ell+1}}Y_{\ell 0}(\theta)=P_{\ell}(\cos \theta) \; , \qquad 
\ee
where $P_{\ell}$ is the Legendre polynomial of order $\ell$, and $f(0)=1$. 
%
%
Then $\mathcal{D}_I  f(\theta) = \left( f'(\theta) , 0 \right)$, and $\mathcal{D}^I  f(\theta)$ has the same components, and,
\be
S_{IJ} = \left[ \begin{array}{cc}
- \frac{1}{2} \ell ( \ell + 1)  f(\theta) - \cot{\theta}  f'(\theta) & 0 \\
0 &  \sin^2{\theta} \left[ \frac{1}{2} \ell ( \ell + 1)  f(\theta) + \cot{\theta} f'(\theta)  \right]
\end{array}  \right] \; .
\ee
We write the linear perturbation as,
\be
\label{eq:pert}
{g}_{ab} = {g}^{({\rm Mink})}_{ab} +  \epsilon \, c_{\ell} \, h_{ab}(t,r,\theta,\phi) 
\ee
where $\epsilon$ is the perturbation parameter and $c_{\ell}$ determines its normalization which we later will choose to depend on $\ell$ in a specific way.
Now we construct these modes from two parts:
\be
h_{ab} = h^{\rm phys}_{ab}  + h^{\rm gauge}_{ab} 
\ee
firstly a `physical' perturbation: 
\be
\label{eq:scalar}
h^{\rm phys}_{ab} = e^{i \omega t} \left[ \begin{array}{ccc}
0 & 0 &  0 \\
0& \delta b(r) f(\theta)  & \delta h_r(r) \mathcal{D}_I  f(\theta) \\
0& \delta h_r(r) \mathcal{D}_I  f(\theta)&  \ \delta h_L(r)  f(\theta) \Omega_{IJ}   +  \delta h_T(r) S_{IJ}\    
\end{array} \right]  + \mathrm{Lie}_\zeta\, g^{\rm (Mink)}_{ab}
\ee
where the vector field $\zeta^a$ is given by,
\be
\begin{split}
    \zeta^t &= - \frac{ i \omega\, e^{i \omega t} f(\theta)}{2} \delta h_T(r) \; , \\
\zeta^r &= \frac{ e^{i \omega t} f(\theta)}{2 }  \left( \delta h_T'(r) - \frac{2}{r} \delta h_T(r) - 2 \delta h_r(r) \right) \; , \\
\zeta^I &= - \frac{ e^{i \omega t} }{2 r^2} \delta h_T(r) \mathcal{D}^I f(\theta) \; . 
\end{split}
\ee
For the convenience of the reader, the explicit components are given in the appendix~\ref{app:smoothness} and we note that the gauge part given by $\zeta^a$ removes terms involving $f'(\theta)$.\footnote{As we discuss later, $h^{\rm phys}_{ab}$ is a smooth tensor at the origin for our perturbation modes when we transform to Cartesian coordinates. Without the gauge part it would not be; the $\zeta^a$ vector field turns out not to be smooth at the origin.}
This is parameterized by the functions $\delta b(r)$ and $\delta h_{r, L, T}(r)$. Imposing the bulk Ricci flatness condition will determine all these, allowing deformations away from the flat spacetime metric, but leaving no residual gauge freedom. Hence we use the terminology `physical' to describe this part.
Secondly we take a ``pure gauge'' piece,
\be
\label{eq:gauge}
h^{\rm gauge}_{ab} = \mathrm{Lie}_\xi\, g^{\rm (Mink)}_{ab}
\ee
with
\begin{equation}
\label{eq:xi}
\xi^t = \frac{1}{2} \delta T(r) f(\theta)e^{i \omega t} \; , \quad
\xi^r =\frac{r}{2}\delta R(r) f(\theta)e^{i \omega t} \; , \quad
\xi^I =\frac{1}{2}\delta Q(r) \mathcal{D}^I f(\theta)e^{i \omega t} \nl
\end{equation}
generated by the vector field $\xi^a$ which we will require to be smooth once transformed to Cartesian coordinates.

Here it is implicit that one must take the real part of these complex perturbations in order to obtain the metric deformation. We emphasize that in the perturbed spacetime the boundary is still at $r=1$, and the initial surface remains at $t=0$. 

Our perturbation is parameterized by the complex functions $\{ \delta b(r) ,  \delta h_r(r) , \delta h_L(r), \delta h_T(r) \}$ and $\{ \delta T(r), \delta R(r), \delta Q(r) \}$, but since the latter three are pure gauge, they do not enter the bulk equations of motion which only involve the first four. The bulk Ricci flatness condition then determines three of these algebraically as,
\begin{equation}
\begin{split}
\delta h_L(r) &= - \frac{1}{2} r^2 \delta b(r)
\\
\delta h_r(r) & = \frac{1}{\ell ( \ell + 1)} \left[ 3 r \delta b(r) + r^2 \delta b'(r) \right]
\\
\delta h_T(r) & = \frac{r^2}{\ell ( \ell + 1) (\ell^2 + \ell - 2)} \left[  ( 6 + \ell + \ell^2 - 2 r^2 \omega^2 ) \delta b(r) + 2 r \delta b'(r) \right]\;,
\end{split}
\label{eq:sol}
\end{equation}
in terms of a ``master variable'' $\delta b$ obeying,
\be
\label{eq:beqn}
\delta b''(r) + \frac{6}{r} \delta b'(r) + \left[  \omega^2 + \frac{6 - \ell( \ell + 1)}{r^2} \right] \delta b(r) = 0\;,
\ee
which is solved in terms of Bessel functions. Imposing the condition of regularity at the origin gives the solution
\be
\label{eq:soln}
\delta b(r) = \frac{\ell^{1/3}}{r^{5/2}} J_{\frac{1}{2} + \ell}(r \omega) \;,
\ee
where we have chosen a particular normalization for $\delta b$ noting that we have encoded the overall freedom of normalizing the perturbation in $c_\ell$ in equation~\eqref{eq:pert}.
This gives a metric perturbation $h_{ab}^{\rm phys}$ that is smooth at $r=0$, as can be checked by converting to Cartesian coordinates. 
Similarly, $h_{ab}^{\rm gauge}$ is smooth at $r=0$ if the vector field $\xi^a$ is smooth at $r=0$. A sufficient condition for this is that,
\be
\label{eq:smoothgauge}
r^{-\ell}\delta T(r), \qquad r^{-\ell} \delta R(r), \qquad r^{-\ell} \delta Q(r)  \qquad  {\rm are\; smooth \; functions \; of \;} r^2 \; .
\ee
In Appendix~\ref{app:smoothness} we give an explicit derivation that shows the smoothness of $h^{\rm phys}_{ab}$ at the origin, and likewise for $\xi^a$.

\subsection{Imposing the boundary conditions}

The above expressions give a solution of the linearized Einstein equation. The ``pure gauge'' part of the perturbation is so far unconstrained. We will exploit this freedom to simplify the solution at the boundary. Specifically we impose
\be
\label{gauge1}
 h_{rt} = h_{r\theta} = h_{r\phi}=0 \qquad \qquad {\rm at} \; r=1\;.
\ee 
This requires that we choose the pure gauge part of the perturbation such that
\be
\label{eq:normal}
 \delta T'(1) =  i \omega \left[ \delta R(1)  -  \frac{4 \omega^2 \left( 4 \delta b(1) +   \delta b'(1) \right) } {\ell (\ell + 1) ( \ell^2 + \ell - 2 ) } \right]
  \; , \quad
  \delta Q'(1) = - \delta R(1)  
 \,.
\ee
Note that $h_{rt}=0$ at $r=1$ implies that $h^{rt}=0$ there, and so the above conditions imply that $\mathrm{d}t$ and $\mathrm{d}r$ are orthogonal at $r=1$. Hence our linear perturbations preserve the property that the initial surface $t=0$ is orthogonal to the boundary $r=1$. In the terminology of \cite{An:2025rlw}, our perturbed metrics all have the same ``corner angle'' (zero) as the background Minkowski spacetime.



The boundary metric, $\gamma_{\mu\nu}$, is now
\begin{equation}
\begin{split}
&\gamma_{tt}=\bar{\gamma}_{tt}-\epsilon \,e^{i \omega t} \left[ i \omega \delta T(1) + \frac{(\lambda_{\ell}+6 - 2 \omega^2 ) \delta b(1) + 2 \delta b'(1)}{\lambda_{\ell} ( \lambda_{\ell} - 2 )}\omega^2   \right]  f(\theta)
\\
&\gamma_{tI}=\bar{\gamma}_{tI}+\epsilon\,\frac{e^{i \omega t}}{2} \left[ i \omega  \delta Q(1) - \delta T(1) \right]  D_I  f(\theta)
\\
&\gamma_{IJ}=\bar{\gamma}_{IJ}+\epsilon\,e^{i \omega t} \left\{ \left[\delta R(1)- \frac{\lambda_{\ell}}{2} \delta Q(1) -  \frac{(\lambda_{\ell}+6) \delta b(1) + 2 \delta b'(1)}{\lambda_{\ell} (\lambda_{\ell} - 2 )} \omega^2  \right]  f(\theta) \Omega_{IJ} + \delta Q(1) S_{IJ} \right\}\,,
\end{split}
\end{equation}
where $\lambda_\ell\equiv\ell(\ell+1)$. Using \eqref{gauge1} the extrinsic curvature of the boundary is
\be
K_{\mu\nu} = \frac{1}{2} \sqrt{g^{rr}} \partial_r g_{\mu\nu} |_{r=1}\;,
\ee
which we may then linearize and trace using the induced metric on the boundary to give at $r = 1$ the quantity appearing in the boundary condition \eqref{bcp}:
\be
\begin{split}
   \frac{ \gamma^p K}{ 2  \bar{\gamma}^p}-1  = \frac{\epsilon}{4} e^{i \omega t} f(\theta) \Bigg\{  &- \omega^2 \frac{  \left[ \lambda_{\ell} - 4 + 8 p - 2 \omega^2 \right]  \delta b'(1) }{\lambda_{\ell} (\lambda_{\ell} - 2 )}   \\&- \omega^2 \frac{   2  \left[ \lambda_{\ell}(1 + 2 p) - 6 + 12 p + \omega^2 ( 4 p - 3) \right] \delta b(1)}{\lambda_{\ell} ( \lambda_{\ell} - 2 )}\\
    &+ 4 i p \omega \delta T(1) - 4 p \lambda_{\ell}\delta Q(1) + [ \lambda_{\ell} - 2 + 8 p - \omega^2 ]\, \delta R(1)   \Bigg\} + \mathcal{O}(\epsilon^2)\;,
\end{split}
\ee
where we have used~\eqref{eq:normal} and the form of the bulk solution. The second boundary condition is that we require the perturbation to the boundary metric does not change the conformal class, i.e., there must exist a function $\psi(t,\theta)$ such that
\be
\gamma_{\mu\nu} = \left[ 1 + \epsilon \, \psi(t,\theta) \right] \bar{\gamma}_{\mu\nu}+ \mathcal{O}(\epsilon^2)
\ee
Using the form of $\delta h_L$ and $\delta h_T$ above in~\eqref{eq:sol} we find that the boundary conditions reduce to a condition at $r = 1$ on the master variable:
\be
\label{eq:bbc}
\frac{\delta b'(1)}{\delta b(1)} = 2  \left[ \frac{6 ( 1 - 6p ) - \lambda_{\ell} (\lambda_{\ell} + 3 + 6 p )  +  (1 +2 \lambda_{\ell}+ 12 p) \omega^2 - \omega^4}{3 \lambda_{\ell} + 24 p - 2(2+\omega^2)} \right]
\ee 
and constraints on the ``pure gauge'' part of the perturbation:
\be
\label{eq:gaugebc}
\delta R(1) &=& - 2  \left[ 1 + \frac{
  2 \omega^2 ( 2 - 12 p + \omega^2 ) - 3 \omega^2 \lambda_{\ell}
}{ \lambda_{\ell} (\lambda_{\ell} - 2 )   } \right] \left[ \frac{\omega^2 \delta b(1) }{3 \lambda_{\ell} - 4 + 24 p  - 2 \omega^2} \right]
\ee
together with,
\be
\label{eq:gaugebc2}
\delta Q(1) &=& 0 \; , \quad \delta T(1) = 0\;,
\ee
with $\psi$ determined by
\be
\psi(t,\theta) =  e^{i \omega t} f(\theta)  \left[ \frac{\omega^2 \delta b(1) }{3 \lambda_{\ell} - 4 + 24 p  - 2 \omega^2} \right] \; .
\ee
Substituting the solution \eqref{eq:soln} for $\delta b(r)$, the boundary condition~\eqref{eq:bbc} reduces to a condition on $\omega$:
\be
\label{eq:omega}
 \frac{J_{\frac{3}{2}+\ell}(\omega)}{J_{\frac{1}{2}+\ell}(\omega) } =  \frac{
(\ell+1)(\ell+2)(2 \ell^2 + \ell - 2 + 12 p) - 2 ( 2 \ell^2 + 3 \ell - 1 + 12 p) \omega^2 + 2 \omega^4
}
{ \omega \left[ 3 \ell ( \ell+1) + 24 p - 2(2 + \omega^2) \right] }\;.
\ee
In summary, given $\ell$ and a corresponding solution $\omega$ of \eqref{eq:omega}, we have constructed a solution of the linearized Einstein equation that satisfies the boundary conditions at $r=1$. The ``pure gauge'' part of the perturbation must satisfy \eqref{eq:normal}, \eqref{eq:gaugebc} and \eqref{eq:gaugebc2} but is not constrained for $r<1$ other than to be smooth at the origin. This arises from the freedom to choose the lapse and shift at $t=0$ which, because of the assumed harmonic time dependence, determines the lapse and shift everywhere. 

\subsection{Unstable modes}

Both the LHS and RHS of \eqref{eq:omega} are odd functions of $\omega$. Hence if $\omega$ satisfies this equation then so does $-\omega$. This symmetry arises from the time-reversal invariance of the background metric, boundary conditions and the form of our Ansatz. Furthermore, if $\omega$ satisfies \eqref{eq:omega} then so does $\bar{\omega}$, so solutions come in complex conjugate pairs. This symmetry arises because if $h_{ab}$ satisfies the equations of motion and boundary conditions then so does its complex conjugate, which has frequency $-\bar{\omega}$. Combining this with the time reversal symmetry gives a mode with frequency $\bar{\omega}$. 

Note that a solution with ${\rm Im}(\omega)<0$ grows exponentially with time so it describes an instability of the flat metric.

For conformal boundary conditions, $p=0$, for each value of $\ell \ge 2$ a conjugate pair of complex solutions of equation \eqref{eq:omega} with ${\rm Re}(\omega)\ge 0$ can be obtained numerically, yielding one unstable mode with ${\rm Im}(\omega)<0$. These unstable solutions for $\omega$ are shown in the first column of Table \ref{tab:omega} for a selection of $\ell$. For other values of $p$, so generalized conformal boundary conditions, we find unstable modes exist for sufficiently large $\ell$. For example, for $p = 1/2$ they exist for $\ell > 12$, and for $p = 1$ they exist for $\ell > 42$. Modes for these values of $p$ are also shown in the table.
\begin{table}
\centering
\begin{tabular}{ |c|c|c|  }
\hline
 \multicolumn{3}{|c|}{$p=0$}\\
\hline
$\ell$&$\omega-\ell$&$\left| \frac{\omega_{\mathrm{asym}}-\omega }{\omega} \right|$\\
\hline
5 & $0.625323 - 0.626343 i$ & $0.0218$\\
\hline
20 & $0.708213 - 0.963743 i$ & $0.00154$\\
\hline
50 & $0.753556 - 1.28344 i$ & $0.000253$\\
\hline
200 & $0.853084 - 2.00662 i$ & $1.60 \times 10^{-5}$\\
\hline
500 & $0.955323 - 2.71061 i$ & $2.56 \times 10^{-6}$\\
\hline
2000 & $1.19395 - 4.28837 i$ & $1.60  \times 10^{-7}$\\
\hline
5000 & $1.42970 - 5.81438 i$ & $2.56 \times 10^{-8}$\\
\hline
 \multicolumn{3}{|c|}{$p=1/2$}\\
\hline
$\ell$&$\omega-\ell$&$\left| \frac{\omega_{\mathrm{asym}}-\omega }{\omega} \right|$\\
\hline
20 & $0.764301 - 0.453929 i$ & $0.00489$\\
\hline
50 & $0.776824 - 0.948394 i$ & $0.000605$\\
\hline
200 & $0.859451 - 1.80645 i$ & $3.41 \times 10^{-5}$\\
\hline
500 & $0.958168 - 2.56514 i$ & $5.33 \times 10^{-6}$\\
\hline
2000 & $1.19491 - 4.19760 i$ & $3.28 \times 10^{-7}$\\
\hline
5000 & $1.43023 - 5.74769 i$ & $5.24 \times 10^{-8}$\\
\hline
 \multicolumn{3}{|c|}{$p=1$}\\
\hline
$\ell$&$\omega-\ell$&$\left| \frac{\omega_{\mathrm{asym}}-\omega }{\omega} \right|$\\
\hline
50 & $0.800117 - 0.388281 i$ & $0.00556$\\
\hline
200 & $0.865829 - 1.58107 i$ & $0.000190$\\
\hline
500 & $0.961018 - 2.41089 i$ & $2.77 \times 10^{-5}$\\
\hline
2000 & $1.19587 - 4.10481 i$ & $1.64 \times 10^{-6}$\\
\hline
5000 & $1.43075 - 5.68021 i$ & $2.58 \times 10^{-7}$\\
\hline
\end{tabular}
\caption{The middle column above shows $\omega - \ell$ for complex solutions of \eqref{eq:omega} with ${\rm Re}(\omega)\ge 0$, and ${\rm Im}(\omega) < 0$ so that they correspond to instabilities. These are shown for a selection of $\ell$ and for several choices of $p$. These solutions are determined numerically. For conformal boundary conditions, $p = 0$, complex solutions exist for all $\ell \ge 2$, whereas for the choices of $p = 1/2$ and $p = 1$, they only exist for $\ell > 12$ and $\ell > 42$, respectively. 
The last column shows the fractional error between this solution and the asymptotic approximation, here denoted $\omega_{\rm asym}$, given by the expression in equation~\eqref{eq:omegaasymp} truncated to the order shown there. We see this approximation converge to the true solution in the large $\ell$ limit, consistent with $\omega - \omega_{\rm asym} \sim O(\ell^{-1})$ as we expect from~\eqref{eq:omegaasymp}. 
}
\label{tab:omega}
\end{table}

\subsection{The instability at large $\ell$}

For $\ell \gg 1$, we can look for solutions of \eqref{eq:omega} with $\omega \sim \ell$ and $\omega - \ell \sim \ell^{1/3}$  using the Olver approximation for the Bessel function in terms of Airy functions \cite{Olver1954,NIST:DLMF}, reviewed in Appendix~\ref{app:Bessel}. In equation~\eqref{eq:BesselApprox} we deduce an asymptotic expansion for $J_{q+\ell}\left[ \ell - \alpha \left( \frac{\ell}{2} \right)^{1/3} \right]$ in the case that $\ell \to \infty$ with $q$ and $\alpha$ fixed, that takes the form of a power series in $\ell^{-1/3}$. Then writing $\omega = \ell - \alpha \left( \frac{\ell}{2} \right)^{1/3}$ and  substituting this into  equation~\eqref{eq:omega}, we find an asymptotic expansion for $\alpha$ in powers of $\ell^{-1/3}$ which yields an expansion for $\omega$ as\footnote{The expansion for $\omega$ in terms of $\ell$ corrects a typo in the $\ell^{-1/3}$ term of Eq.~(4.50) in \cite{Liu:2024ymn}.
},
\be\label{eq:omegaasymp}
\omega = \ell - \Gamma \left( \frac{\ell}{2} \right)^{1/3} + \frac{1}{2} + \frac{13 - 120 p - 187 \Gamma^3 + 48 \Gamma^6 }{20 \Gamma (7 + 16 \Gamma^3)} \left( \frac{2}{\ell} \right)^{1/3} - \frac{\Gamma}{12} \left( \frac{2}{\ell} \right)^{2/3} + \mathcal{O}\left(\ell^{-1} \right)\,.
\ee
where $\Gamma$ is a solution of
\be
\label{eq:gammaeqn}
\mathrm{Ai}'(\Gamma) = 4 \Gamma^2 \mathrm{Ai}(\Gamma)\,.
\ee
A particular solution of this can be found numerically:
\be
\label{eq:Gammaval}
\Gamma \equiv \Gamma_r + i \Gamma_i \; , \qquad \Gamma_r = - 0.067424262 \; , \quad \Gamma_i = 0.427904905 \;,
\ee
the key point being that this is complex and leads to unstable exponential growth in time due to the harmonic time dependence $e^{i \omega t}$. 
The conjugate $\bar{\Gamma} = \Gamma_r - i \Gamma_i$ yields another solution to~\eqref{eq:gammaeqn}, but is not of interest here as it gives a perturbation that decays in time. These appear to be the only complex solutions although we note there are infinitely many real solutions where $\Gamma < 0$.
Truncating~\eqref{eq:omegaasymp} at order $\ell^{-2/3}$ gives the approximation $\omega_{\rm asym}$ shown in the second column of table \ref{tab:omega}, which is in increasingly good agreement with our numerical results as $\ell$ increases.

While we have numerically constructed $\omega$ corresponding to instabilities for quite large $\ell$, and have an asymptotic approximation of these solutions, this does not prove such modes exist for arbitrarily large $\ell$. In fact we may straightforwardly prove that given the solution~\eqref{eq:Gammaval} for $\Gamma$ of the condition~\eqref{eq:gammaeqn} then for sufficient large $\ell$ we are guaranteed solutions of~\eqref{eq:omega} where $\omega$ is approximated by the asymptotic expansion~\eqref{eq:omegaasymp}. The argument is brief and given in Appendix~\ref{app:omega}.

At large $\ell$ the imaginary part of $\Gamma$ leads to an unstable mode growing exponentially in time: 
\be
\left| e^{i \omega t} \right| \simeq e^{\Gamma_i \left( \frac{\ell}{2} \right)^{1/3} t} \; .
\ee
The growth rate of the instability increases unboundedly with $\ell$, just as in the initial value problem for the Laplace equation, which suggests that the initial-boundary value problem is not well-posed. We will show below that this is indeed the case. 

\section{Ill-posedness}
\label{sec:illposed}

Having seen the instability of linear perturbations of the cavity, we will now use these to argue that they form an obstruction to well-posedness of the linear IBVP. In the first two subsections we will give the proof of ill-posedness given certain conditions hold for our unstable modes. In the remainder of the section we prove that those conditions do indeed hold.

\subsection{Gauge invariant quantities}

Discussions of well-posedness in General Relativity are complicated by the need to pick a gauge. For the standard initial value problem, some choices of gauge lead to a well-posed problem whereas others do not. Fortunately in our problem there exist gauge-invariant quantities, which we will describe in this section. If there exists a gauge in which the problem is well-posed then, by working in this gauge, the gauge-invariant quantities must depend continuously on the initial (and boundary) data. Conversely, if we can find a family of solutions that demonstrates that these quantities do not possess such continuous dependence then we have demonstrated that there cannot exist a gauge in which the problem is well-posed. 

Since we are perturbing around a flat background, the simplest gauge invariant quantity is the linear perturbation in the Riemann tensor. Gauge-invariance implies that this does not depend on $h^{\rm gauge}_{ab}$ and is therefore determined entirely by $\delta b(r)$ and $\delta b'(r)$ where higher derivatives of $\delta b$ can be eliminated using its equation of motion. A particularly simple component is $R_{trtr}$.
We denote the linear perturbation to this as $\mathcal{R}$  and then,
\be\label{eq:gaugeinvR}
R_{trtr}  = \epsilon\, \mathcal{R} + \mathcal{O}(\epsilon^2) \; , \qquad
\mathcal{R}  =  \frac{c_\ell e^{i \omega t} \omega^2}{2} f(\theta) \delta b(r) 
\ee
which is then gauge invariant.
Another natural quantity to consider is the Ricci scalar of the (Lorentzian) boundary geometry. This is gauge invariant since the unperturbed Ricci scalar is constant so a gauge transformation leaves it invariant. It can be expressed in terms of $\mathcal{R}$ at $r = 1$ as
\be
R_\gamma = 2 \left[1 - \epsilon\, \frac{2 (\ell^2 + \ell -1 - \omega^2)}{3\ell (\ell+1) + 24 p - 2(2 + \omega^2)} \left( \left. \mathcal{R} \right|_{r=1} \right) + \mathcal{O}(\epsilon^2) \right]\;.
\ee

\subsection{Proof of ill-posedness}
\label{sec:proof}

We will follow the logic of showing ill-posedness of the Laplace equation  as a Cauchy problem, as in section \ref{sec:laplace}. All of our modes satisfy the same boundary conditions at $r=1$ and the same corner condition at $r=1$, $t=0$. However, they have different initial data. We will now argue that we can construct a sequence of these modes demonstrating that the gauge-invariant quantity $\mathcal{R}$ does not depend continuously on the initial data, thus proving that the problem is not well-posed in any gauge. 

The initial data for linearized gravity are the (linearized) induced metric and extrinsic curvature of the $t=0$ surface. For our modes, these quantities are determined by $\delta b(r)$ and the ``gauge functions'' $\{ \delta T(r), \delta R(r), \delta Q(r) \}$. We have not yet fixed the latter.
These are constrained by the cavity boundary conditions~\eqref{eq:gaugebc} and \eqref{eq:gaugebc2} which fix the values of these quantities at $r = 1$ in terms of the function $\delta b(r)$ and its derivative there. Furthermore the condition~\eqref{eq:normal} determines derivatives of two of these, $\delta T'(1)$ and $\delta Q'(1)$ again in terms of $\delta b(1)$ and $\delta b'(1)$. We note that $\delta R'(1)$ is unconstrained. In addition we must satisfy the smoothness conditions at the origin~\eqref{eq:smoothgauge}. 
Let us define the function,
\be
\Theta(r) = - \frac{r^\ell}{2} \left( 1 - r^2 \right)
\ee
which, for integer $\ell$, is smooth on the interval $[0,1]$. Then at the cavity boundary we have $\Theta(1) = 0$ and $\Theta'(1) = 1$. 
We now make an explicit choice for the gauge functions as, 
\be
\label{eq:gaugechoice}
&& \delta T(r) = \Theta(r) \delta T'(1) \; , \quad \delta R(r) = r \Theta'(r) \delta R(1)  \; , \quad \delta Q(r) =  \Theta(r) \delta Q'(1) 
\ee
and we take $\{ \delta R(1), \delta T'(1), \delta Q'(1) \}$ to be determined from $\delta b(1)$ and $\delta b'(1)$ as in~\eqref{eq:normal} and~\eqref{eq:gaugebc}. Then $\delta T(1), \delta Q(1)$ vanish, so are consistent with~\eqref{eq:gaugebc2}. By construction these gauge functions satisfy the conditions~\eqref{eq:smoothgauge} for smoothness of $h^{\rm gauge}_{ab}$ and satisfy all the boundary conditions.
We note that for the choice of gauge functions above, 
transforming to Cartesian coordinates $x^{a'} = (t,x,y,z)$ for the background Minkowski spacetime in the usual manner (see Appendix~\ref{app:smoothness}), then the components of the perturbation $h_{a'b'} = h^{\rm phys}_{a'b'} + h^{\rm gauge}_{a'b'}$ in these coordinates are
 not just smooth,  but are are in fact analytic functions.

We need to understand how our initial data behaves as $\ell \rightarrow \infty$. We are going to show below that for integer $n$ and $m$ then $r^{-m} \partial_r^n \delta b(r)$ is uniformly (in $r$) bounded by a ($n$-dependent) power of $\ell$ for $r \in [0,1]$ as $\ell \to \infty$. This implies that taking $n$ derivatives of any of the functions in \eqref{eq:gaugechoice} gives a function that is also uniformly bounded by a ($n$-dependent) power of $\ell$ (since these depend on the values of $\delta b(1)$ and $\delta b'(1)$). 
Taking any number of spatial derivatives of the initial perturbation $h_{a'b'}|_{t=0}$ in Cartesian coordinates gives derivatives of $\delta b(r)$ and the gauge functions, and derivatives of $f(\theta)$ (which are also uniformly bounded by powers of $\ell$), all multiplied by inverse powers of $r$. Hence they are uniformly bounded by a power of $\ell$ as $\ell \rightarrow \infty$.
Finally, $\partial_t^n h_{a'b'}|_{t=0} = (i\omega)^n h_{a'b'}|_{t=0}$, which is bounded by a power of $\ell$ since $\omega = \mathcal{O}(\ell)$. These results imply that the initial data for $h_{a'b'}$ is uniformly bounded by a power of $\ell$ as $\ell \rightarrow \infty$. Similarly if we consider any number of derivatives of this data then this is also uniformly bounded by a power of $\ell$ as $\ell \rightarrow \infty$. Similar statements apply to the linearized metric and extrinsic curvature of the $t=0$ surface.

We now make the following choice for the normalization constant appearing in \eqref{eq:pert} 
\be\label{eq:omegachoice}
 c_\ell = \exp(- \ell^{1/6})\;.
\ee
As $\ell \rightarrow \infty$, the exponential decay of $c_{\ell}$ beats the power law growth of any derivative of our initial data. So all $C^k$ norms of our initial data tend to zero as $\ell \rightarrow \infty$. Hence we have a sequence of initial data that tends to zero in $C^\infty$ as $\ell \rightarrow \infty$. If the problem were well-posed then the resulting sequence of solutions should also converge to zero. 
However, for $t>0$ the gauge invariant quantity $\mathcal{R} \propto c_\ell e^{i\omega t} \delta b(r)f(\theta)$.
We will show that the value of $\mathcal{R}$ at the boundary evaluated at $\theta=0$ has a magnitude that is bounded below as,
\be
| \mathcal{R}_{r=1,\theta=0} | > k\, \ell^{2} e^{-\ell^{1/6}} e^{\Gamma_i  \left( \frac{\ell}{2}\right) ^{1/3} t} 
\ee
for a constant $k > 0$. This diverges as $\ell \rightarrow \infty$ for any $t>0$. Hence our sequence of solutions does not converge in $C^\infty$. 
Since $\mathcal{R}$ is gauge-invariant, this proves that there cannot exist a gauge in which the solution depends continuously on the initial data.
This argument proves that the map from initial data to solutions (if it exists) is not continuous at the point corresponding to trivial initial data. By adding our sequence of solutions to any other solution as in section \ref{sec:laplace} demonstrates a lack of continuity at an arbitrary point in the space of initial data.

\subsection{Bounding $\delta b$ and its derivatives as $\ell \to \infty$}
\label{sec:bounds}

\begin{figure}
    \centering
    \includegraphics[width=0.45\linewidth]{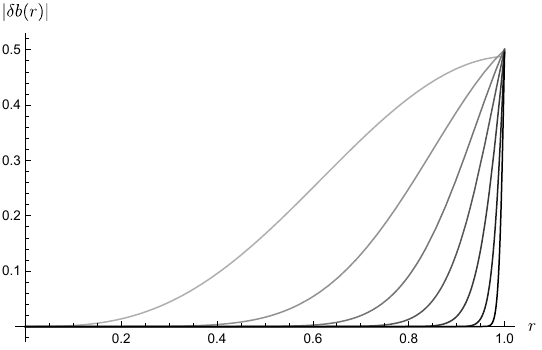} \hspace{0.5cm} \includegraphics[width=0.45\linewidth]{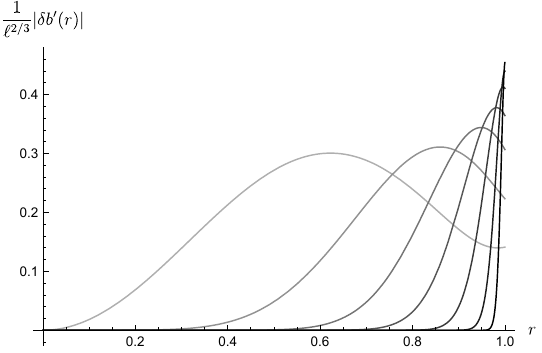} \\
    \includegraphics[width=0.45\linewidth]{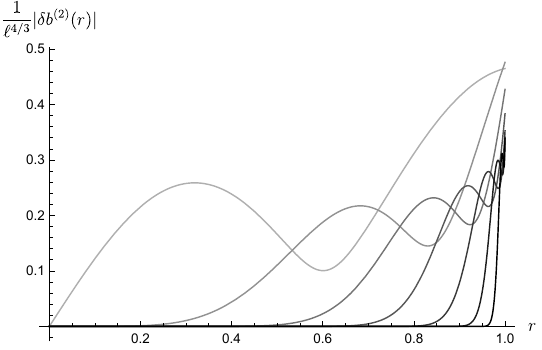} \hspace{0.5cm} \includegraphics[width=0.45\linewidth]{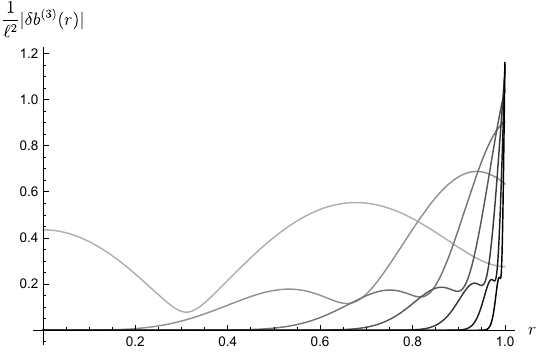}
    \caption{Plot of the magnitude of $\partial^n_r \delta b(r)$ for $n = 0,1,2$ and $3$, scaled by $\ell^{- \frac{2 n}{3}}$ for the case of conformal boundary conditions, $p=0$. These are plotted against $r$ over the whole domain $[0,1]$ for $\ell = 5, 10, 20, 40, 100, 300$ and $1000$ (light gray to black for increasing $\ell$). As $\ell$ becomes large, the curves become increasingly localized near the boundary $r = 1$. The plots suggest that $\ell^{- \frac{2 n}{3}} \partial^n_r \delta b(r)$ is $\mathcal{O}(1)$ as $\ell \to \infty$.  }

    \label{fig:b_and_derivs}
\end{figure}

We  now consider  the behaviour of $\delta b(r) = \ell^{1/3}J_{\frac{1}{2} + \ell}(r \omega)/r^{5/2}$ as $\ell \rightarrow \infty$. 
Our goal is to show that, for any $n$ and $m$, as $\ell \rightarrow \infty$, $\left| r^{-m} \partial^n_r \delta b(r) \right|$ is uniformly bounded by a power of $\ell$, justifying our claim above.
More formally, for any given integers $m$ and $n$, we wish to show that there exists $s\in\mathbb{R}$ and constants $C_1 > 0$ and $\ell_0$ such that for $\ell>\ell_0$ we have
\be
\label{eq:bound}
\left| \frac{1}{r^m} \partial^n_r \delta b(r) \right| < C_1 \ell^s
\ee
for all $r \in [0,1]$. Before we do this we may gain some intuition by plotting the magnitude of $\delta b(r)$ and its first few derivatives over our domain for a sequence of increasing $\ell$. We present these plots in figure~\ref{fig:b_and_derivs} where we take $m = 0$ and we choose $p=0$, i.e., Anderson's conformal boundary conditions. 
These plots suggest that $ \ell^{- \frac{2 n}{3}} \sup \left| \partial^n_r \delta b(r) \right|$ is $\mathcal{O}(1)$ as $\ell \to \infty$. In the rest of this section we will show the weaker result that $ \ell^{- n} \sup \left| \partial^n_r \delta b(r) \right|$ is $\mathcal{O}(1)$ in this limit.

\begin{figure}
    \centering
    \includegraphics[width=0.45\linewidth]{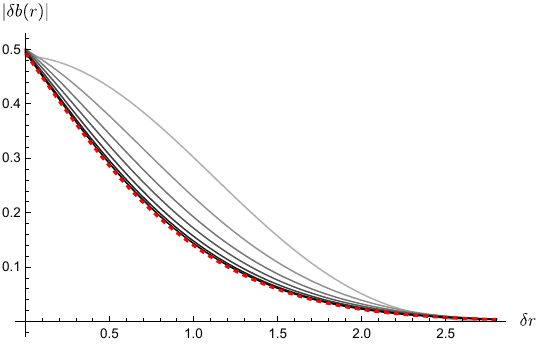} \hspace{0.5cm} \includegraphics[width=0.45\linewidth]{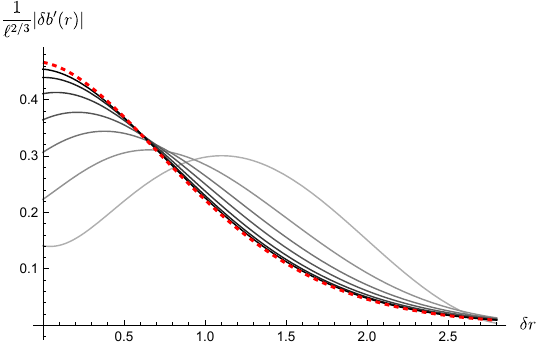} \\
    \includegraphics[width=0.45\linewidth]{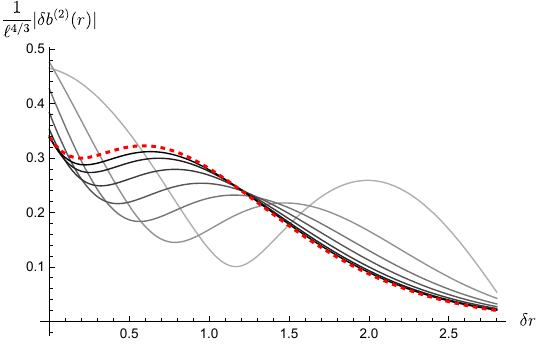} \hspace{0.5cm} \includegraphics[width=0.45\linewidth]{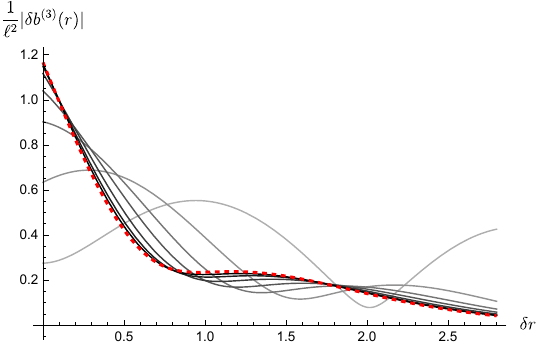}
    \caption{Plot of the same quantity as in figure~\ref{fig:b_and_derivs}, now against $\delta r\sim \mathcal{O}(1)$ so that we focus on the near boundary region. The plots appear to asymptote to limiting curves (the red dashed lines) as $\ell \rightarrow \infty$. This limiting behaviour is described by equation~\eqref{eq:asymb} below.}
    \label{fig:b_and_derivs_nearboundary}
\end{figure}

For large $\ell$ we see from the plots that $\delta b(r)$ becomes increasingly localized near the boundary. It is convenient to introduce the ``near boundary'' coordinate $\delta r$ as,
\be\label{eq:defdeltar}
r = 1 - \frac{\delta r}{\ell^{2/3}} \; .
\ee
Then $\delta r = 0$ is the boundary $r = 1$, and we define the ``near boundary region'' as the region with $\delta r \sim \mathcal{O}(1)$ as we take $\ell \to \infty$. In figure~\ref{fig:b_and_derivs_nearboundary} we plot the same functions as in the previous figure~\ref{fig:b_and_derivs}, but now against the near boundary coordinate.

For large $\ell$ we see $\delta b(r)$ localizes in the near boundary region. Defining $\alpha(\delta r)$ from,
\be
\label{eq:alphar}
r \omega 
= \ell - \alpha( \delta r) \left( \frac{\ell}{2} \right)^{1/3}
\; , \qquad \mathrm{so\;that} \qquad
\alpha( \delta r) = \Gamma + 2^{1/3} \delta r + \mathcal{O}\left( \ell^{-1/3} \right) \;,\nl
\ee
then from the asymptotic approximation in Appendix~\ref{app:Bessel}  
we find that the leading behaviour of $\delta b(r)$ in the near boundary region is, 
\be
\delta b(r) \simeq 2^{1/3} \mathrm{Ai}\left[ \alpha(\delta r) \right] \simeq 2^{1/3} \mathrm{Ai}(\Gamma + 2^{1/3} \delta r) 
\ee
as $\ell \to \infty$.  Moving away from the boundary, the Airy function decays exponentially for $\delta r \gg 1$. 
Hence $\delta b$ localizes in the near-boundary region which has width $\sim \ell^{-2/3}$ as $\ell \to \infty$.
Taking $n$ $r$-derivatives then yields
\be
\label{eq:asymb}
\partial^n_r \delta b(r) \simeq (-1)^n 2^{\frac{n+1}{3}} \ell^{\frac{2n}{3}} \mathrm{Ai}^{(n)}(\Gamma + 2^{1/3} \delta r) 
\ee
and so we see $\partial^n_r \delta b(r) \sim \mathcal{O}\left(\ell^{\frac{2n}{3}}\right)$ in the near boundary region. 

These asymptotic expressions are shown in figure~\ref{fig:b_and_derivs_nearboundary} and indeed we see convergence towards them as $\ell \rightarrow \infty$. These results, and the figures, suggest that the bound in~\ref{eq:bound} holds choosing $s = 2 n/3$. However, to prove this, we must show that this bound holds on the whole domain $r \in [0,1]$, not just in the near-boundary region. We will now do this, although we will actually only prove a weaker bound, with $s=n$, which is good enough for our purposes.

In order to prove the result in~\eqref{eq:bound} we will start by considering the function $u(r)$ defined as, 
\be
\label{eq:fnu}
u(r) = \frac{\ell^{1/3}}{r^{q + a}} J_{q + \ell}(r \omega)
\ee
on our domain $r \in [0,1]$ with $q$ real, and $a, \ell$ integers, and $\omega$ given as a complex solution to~\eqref{eq:omega}, which at large $\ell$ has the behaviour~\eqref{eq:omegaasymp}. We further restrict to $\ell > a$, and then the function $u(r)$ is smooth on our domain. In particular as $r \to 0$  we have $u(r) \sim r^{\ell - a}$ so $u(0) = 0$.  We note that $\delta b(r)$ is of this form with $q = 1/2$ and $a = 2$. However, we will use this more general form to deduce the bound on derivatives of $\delta b(r)$. 
In Appendix~\ref{app:proofs} we prove the following result:

\begin{lemma}
\label{eq:lemma}
There exists $L$ and $C > 0$ (independent of $\ell$) such that if $\ell>L$ then  $| u(r) | < C$ for any $r \in [0, 1]$.
\end{lemma}

\noindent
We see this proves the bound on $\delta b(r)$: \eqref{eq:bound} holds with $s=n=0$. Now in order to extend this to derivatives of $\delta b(r)$ we consider the function $v(r)$ defined as,
\be\label{eq:vofr}
v(r) = \frac{\ell^{1/3}}{r^{\frac{1}{2} + b}} \partial_r^N J_{\frac{1}{2} + \ell}(r \omega) 
\ee
for integer $b$. We now make use of the fact that we may write,
\be
v(r) = \frac{\omega^N}{2^N}  \sum_{m = 0}^N (-1)^m C^N_m \frac{\ell^{1/3} J_{ \frac{1}{2} + 2 m - N  + \ell}(r \omega)}{r^{\frac{1}{2} + b}}\;,
\ee
where $C^N_m = N!/(m! (N-m)!)$ are the binomial coefficients.
Now  $\omega^N = \ell^N \cdot\left[ 1 + \mathcal{O}(\ell^{-2/3}) \right]$, and each summand has the form of $u(r)$ above, so $\ell^{1/3} J_{q + \ell}(r \omega)/r^{q + a}$ with $b-N \le {a} \le b+ N$,  so is bounded as above provided that $\ell > b + N$.  Since we have
 a finite sum of these terms we may find $L'$ and $C_2$ such that for $\ell>L'$ we have
\be
\left| v(r) \right| < C_2 \ell^{N}
\ee
over the whole domain $[0,1]$.
Finally we return to derivatives of our function, $\delta b(r) = \ell^{1/3} J_{\frac{1}{2} + \ell}(r \omega)/r^{5/2}$, multiplied by inverse powers of $r$. Its
$n$-th $r$-derivative, $\partial^n_r \delta b(r)$, can be written as a finite sum of terms of the form of $v(r)$ above in~\eqref{eq:vofr}, with the maximal $N$ being $n$.
Hence for any integers $m$ and $n$, there exists $C_3 > 0$ and $L''$ such that for $\ell>L''$ we have
\be
\label{eq:derivdeltabbound}
\left| \frac{1}{r^m} \partial^n_r \delta b(r) \right| < C_3 \ell^{n}
\ee
for all $r \in [0,1]$. This proves that~\eqref{eq:bound} holds with $s = n$, although as we have noted above, this bound is probably not optimal. However, this bound is sufficient for our purposes.

We have concluded that the $n$-th derivative of $\delta b(r)$ is indeed uniformly bounded in magnitude by a power of $\ell$ on our domain as claimed in Section~\ref{sec:proof}. 
This remains true if $\delta b^{(n)}(r)$ is multiplied by integer powers of $r$, essentially since $\delta b(r)$ localizes near the boundary where $r \simeq 1$.
Then since $\omega = \ell + \mathcal{O}(\ell^{1/3})$, for our choice of $c_{\ell} = e^{- \ell^{1/6}}$ then we have 
$\Big\lVert c_{\ell}\, \frac{\omega^n \, \ell^q}{r^m} \delta b(r) \Big\rVert_{C^k} \to 0$ where the norm is taken over $r \in [0,1]$.

Now consider the physical metric perturbation in Cartesian coordinates $x^{a'} = (t, x, y, z)$. 
The expressions for these components $h^{\rm phys}_{a'b'}$ are given in equation~\eqref{eq:metcart} in Appendix~\ref{app:smoothness} and we see 
that up to factors of the Cartesian coordinates, these are given by the functions $\Phi$ and $\Pi$ in equation~\eqref{eq:phipi}. These are, in turn, products of powers of $r$, $\delta b(r)$ and its first derivative $\delta b'(r)$, the angular dependence $f(\theta)$, and the harmonic time dependence.
Now consider spatial Cartesian derivatives of these functions $\Phi$ and $\Psi$.
We may ignore the time dependence for the moment as we are interested  in spatial derivatives, and then these take the form $F(r,\theta) = u(r) f(\theta)$ so that for $\Phi$, $u(r) = r^2 \delta b(r)$, and in the case it is $\Pi$, instead $u(r) = r^3 \delta b'(r)$. 
Now consider $n \ge 1$ Cartesian derivatives of $F$. This may be written as,
\be
\partial_{i_1} \ldots \partial_{i_n} F = \sum_{a,b = 0}^{a+b \le n} \frac{A^{a,b}_{i_1 \ldots i_n}(\theta, \phi)}{r^{n-a}} u^{(a)}(r) P^{(b)}_\ell(\cos(\theta))
\ee
where crucially the coefficients $A^{a,b}_{i_1 \ldots i_n}(\theta, \phi)$ are smooth functions of $\theta$ and $\phi$, and we have used the explicit form of $f(\theta) = P_\ell(\cos(\theta))$.
Given our bound above in~\eqref{eq:derivdeltabbound}, then for both the case of $\Phi$ and $\Psi$ we have that for a given $M$ there exists $C>0$ such that for sufficiently large $\ell$ the function $u(r)$ and its first $M$ derivatives are all bounded as,
\be
\left| \frac{1}{r^{M - m}} u^{(m)}(r) \right| \le C \ell^{m+1}
\ee
over $r$ in $[0,1]$ for all $0 \le m \le M$. 
Also  there exists $C'>0$ such that for all $0 \le m \le M$ then $P^{(m)}_\ell(\cos(\theta))$ is uniformly bounded for all $\theta$ as $\left| P^{(m)}_\ell(\cos(\theta)) \right| \le C' \ell^{2 m}$ as $\ell \to \infty$.
\footnote{
Writing $z = \cos{\theta}$, so that we are concerned with the interval $z \in [-1,1]$, then $P^{(m)}_\ell(z)$ may be written in terms of Jacobi polynomials as,
$
\frac{d^m}{d z^m} P_\ell(z) = \frac{\Gamma\left[ \ell+ m + 1 \right]}{2^m \Gamma\left[ \ell+ 1 \right]} P^{(m,m)}_{\ell-m}(z)
$
(using \cite{NIST:DLMF} \S18.9)
and further for $z \in [0,1]$ then 
$\left| P^{(m,m)}_{\ell-m}(z) \right| \le \left| P^{(m,m)}_{\ell-m}(\pm 1) \right| = \left( \begin{array}{c} \ell \\ \ell - m \end{array} \right)$ (from \cite{NIST:DLMF} \S18.14). 
Hence  $\left| P^{(m,m)}_{\ell-m}(z) \right| = (\ell^m)$ at fixed $m$ and large $\ell$ and  using that $ \frac{\Gamma\left[ \ell+ m + 1 \right]}{2^m \Gamma\left[ \ell+ 1 \right]}  = O(\ell^m)$ we have,
$\left| \frac{d^m}{d z^m} P_\ell(z) \right| = O(\ell^{2 m})$.
}
Thus we see that as $\ell \to \infty$  for fixed derivative order $n \ge 1$,
\be
\left| \partial_{i_1} \ldots \partial_{i_n} F \right| &\le& \sum_{a,b = 0}^{a+b \le n} 
\left| A^{a,b}_{i_1 \ldots i_n} \right| \left| \frac{u^{(a)}(r)}{r^{n-a}} \right| \left| P^{(b)}_\ell(\cos(\theta)) \right| \le C \,C' \sum_{a,b = 0}^{a+b \le n} 
\left| A^{a,b}_{i_1 \ldots i_n} \right| \ell^{a + 2 b + 1}
= O\left( \ell^{2 n + 1} \right) \; . \nl
\ee
Hence Cartesian spatial derivatives of $\Phi$ and $\Pi$ are bounded by powers of $\ell$ as $\ell \to \infty$.
It follows that the Cartesian spatial derivatives of the Cartesian components $h^{\rm phys}_{a'b'}$ are bounded by powers of $\ell$ as $\ell \to \infty$.

Due to the harmonic time dependence of the functions $\Phi$ and $\Pi$ then $\partial_t^{m} \left( \partial_{i_1} \ldots \partial_{i_n} h^{\rm phys}_{a'b'} \right) = ( i \omega )^{m} \left( \partial_{i_1} \ldots \partial_{i_n} h^{\rm phys}_{a'b'} \right)$, and since $\omega = O(\ell)$ additional time derivatives leave $\partial_{i_1} \ldots \partial_{i_n} h^{\rm phys}_{a'b'}$ polynomially bounded by $\ell$. Thus finally  
we conclude that,
\be
\Big\lVert \partial^N_t \left( c_{\ell} h^{\rm phys}_{a'b'} \right)\Big|_{t=0} \Big\rVert_{C^k} \to 0 \;,\quad \mathrm{as} \quad \ell\to \infty
\ee
where this norm is taken over the full interior of the cavity, $x^2+y^2+z^2 \le 1$.
Now $h^{\rm gauge}_{a'b'}$ is given in terms of the vector field $\xi^{a'}$ whose form in Cartesian coordinates is given explicitly in equations~\eqref{eq:gaugechoice} and~\eqref{eq:xicart}.  A similar argument to that above shows that mixed time and Cartesian spatial derivatives of $\xi^{a'}$ are bounded polynomially in $\ell$ at large $\ell$, and hence,
\be
\Big\lVert \partial^N_t \left( c_{\ell} h^{\rm gauge}_{a'b'} \right)\Big|_{t=0} \Big\rVert_{C^k} \to 0 \;,\quad \mathrm{as} \quad \ell\to \infty\; .
\ee
Thus we see $\Big\lVert \partial^N_t \left( c_{\ell} (h^{\rm phys}_{a'b'} + h^{\rm gauge}_{a'b'}) \right)\Big|_{t=0} \Big\rVert_{C^k} \to 0$ in the limit $\ell\to 0$ so that the initial perturbation converges to zero in the $C^k$ norm for Cartesian coordinates over the interior of the cavity.  Hence, as discussed in Section~\ref{sec:proof}, this implies that as $\ell \to \infty$ we have a sequence of smooth solutions,  $c_{\ell} (h^{\rm phys}_{ab} + h^{\rm gauge}_{ab})$, all with the same (zero) initial corner angle, whose initial data converges in $C^k$ to trivial initial data as $\ell \rightarrow \infty$.

\subsection{Lack of continuity of solutions on initial data}

Trivial initial data with zero corner angle admits the zero solution for the linear gravitational dynamics.
However we now show that for $t > 0$ the solutions in our sequence do not converge to this zero solution. Indeed the sequence does not even have a limit. We consider the gauge invariant $\mathcal{R}$ in~\eqref{eq:gaugeinvR}. For any time $t > 0$ then
 using the asymptotic expansion in Appendix~\ref{app:Bessel}  
we find in the near boundary region the asymptotic behaviour, 
\be
\mathcal{R} &=& \ell^2 e^{- \ell^{1/6}} e^{i \omega t}  2^{-2/3}  \mathrm{Ai}(\Gamma + 2^{1/3} \delta r ) f(\theta) + \mathcal{O}\left( \frac{1}{\ell} \right)
\; .
\ee
Using that $\mathrm{Im}(\omega) =  - \Gamma_i  \left( \frac{\ell}{2}\right)^{1/3} + \mathcal{O}(\ell^{-1/3})$ then
 for sufficiently large $\ell$ we have that at the boundary $\delta r = 0$ then $\mathcal{R}$ evaluated at $\theta=0$ is bounded in magnitude from below as, 
\begin{equation}\label{eq:boundR}
    |\mathcal{R}_{r=1,\theta=0}| > \ell^2 e^{- \ell^{1/6}} e^{\Gamma_i  \left( \frac{\ell}{2}\right) ^{1/3} t}  \left| \frac{ \mathrm{Ai}(\Gamma) }{2} \right|
\end{equation}
%
%
as claimed earlier in~\ref{sec:proof}.
Now recalling that $\Gamma_i > 0$, then $\ell^2 e^{- \ell^{1/6}} e^{\Gamma_i  \left( \frac{\ell}{2}\right) ^{1/3} t} \to \infty$ as $\ell \to \infty$. Hence we see that for any fixed $t>0$ this boundary gauge invariant $\mathcal{R}_{r=1,\theta=0}$ diverges as $\ell \to \infty$ as claimed in Section~\ref{sec:proof}. 
Indeed at $\theta = 0$, $\mathcal{R}$ diverges everywhere in the near boundary region, too. Hence, as discussed above, our sequence does not limit to the zero solution, showing that solutions do not continuously depend on initial data, and therefore that the linear problem cannot be well-posed in any gauge.

\section{Exterior geometry}\label{sec:outside}

So far, we have focused our attention on solving for perturbations in the interior of a spherical cavity that are regular at $r=0$ and satisfy the boundary conditions given in Eq.~(\ref{bcp}). One might wonder whether the same instability also occurs in the exterior of a spherical boundary. Taking the same generalized conformal boundary conditions at the timelike spherical boundary at $r = 1$, we then consider initial data at $t =0$ extending out to spatial infinity $r \to \infty$. 
We may repeat the analysis in Section~\ref{sec:lingrav}, finding that imposing the boundary conditions gives the same conditions as in equation~\eqref{eq:bbc}. Now in the exterior of the boundary the relevant solution of~(\ref{eq:beqn}) that we choose for $\delta b(r)$ is, 
\begin{equation}
\delta b(r) = \frac{\ell^{1/3}}{r^{5/2}} H_{\frac{1}{2} + \ell}(r \omega) \;,
\end{equation}
where $H_{\nu}(x)$ is an order-$\nu$ Hankel function of the second kind. At large $r$ this goes as $e^{i \omega t} \delta b(r) \sim e^{i \omega (t - r)}$ so corresponds to taking the solution with outgoing radiation at future null infinity.
Then performing a similar expansion to the one we outlined for the interior solution,  using the expansion of the Hankel function given in Appendix~\ref{app:Bessel}, now gives
\begin{equation}
\omega = \ell+e^{-\frac{i \pi }{3}}\Gamma\left(\frac{\ell}{2}\right)^{1/3}+\frac{1}{2}+\frac{13-120 p-187 \Gamma ^3+48 \Gamma ^6}{20 \Gamma  \left(7+16 \Gamma ^3\right)}e^{-\frac{2i \pi}{3}}\left(\frac{2}{\ell}\right)^{1/3}+e^{-\frac{i\pi}{3}}\frac{\Gamma}{12}\left(\frac{2}{\ell}\right)^{2/3}+\mathcal{O}(\ell^{-1})\, ,
\end{equation}
where again $\Gamma$ solves~\eqref{eq:gammaeqn} and we now take the conjugate solution $\bar{\Gamma} = \Gamma_r - i \Gamma_i$ to that for the interior case, so $\Gamma_{r,i}$ are given in Eq.~(\ref{eq:Gammaval}) as previously. 
Then we see that $\delta b(r) \sim e^{-(\Gamma_i/2+\sqrt{3}\Gamma_r/2)(\ell/2)^{1/3} r}$ and so decays\footnote{Note that from~\eqref{eq:Gammaval}, $\Gamma_i/2+\sqrt{3}\Gamma_r/2\sim 0.156>0$. } at large $r$. Again as $\ell \to \infty$ we find these modes localize in a near boundary region of width $\sim \ell^{-2/3}$, and that the metric perturbation and its time derivative both vanish at $t = 0$ in this limit.
For $t > 0$ we have that $\left| e^{i \omega t} \right| \sim \exp[(\Gamma_i/2+\sqrt{3}\Gamma_r/2)(\ell/2)^{1/3}t ]$ and so the metric perturbation does not converge to zero in the near boundary region.
This implies that the linear problem is not well-posed in the exterior of a spherical boundary.

\section{Cylindrical cavity}\label{sec:cylindrical}

We have seen that the IBVP for linearized gravity with (generalized) conformal boundary conditions is not well-posed either inside or outside a spherical boundary. The failure of well-posedness is demonstrated by the families of exponentially growing modes constructed above. 
Our calculations above took the cavity to have unit radius. Scaling it to have radius $R$, then the instability has the behaviour,
\be \label{eq:omegaofR}
\omega \simeq \frac{\ell}{R} - \frac{\Gamma}{R^{2/3}} \left( \frac{\ell}{2 R} \right)^{1/3} 
\ee
where $\Gamma$ is the same as above in equation~\eqref{eq:Gammaval}.
Taking a larger cavity and fixing $R/\ell$, the physical spatial length scale of the instability since it localizes at the boundary, we see the instability time scale becomes slower, going as $R^{-2/3}$ as we take $R \to \infty$. Thus the exponential growing instability `switches off' on fixed scales as the boundary becomes less curved. Hence we see that this instability is a subtle sub-leading effect that is due to the curvature of the boundary.
A sphere has both intrinsic and extrinsic curvature. To better understand which type of curvature drives the instability in this section we will consider a cylindrical boundary. This has extrinsic, but not intrinsic, curvature. Nevertheless, we will see that there is a family of exponentially growing modes very similar to those constructed above. (In fact we will see that the growth rate of modes with high (angular) frequency is governed by exactly the same quantity \ref{eq:Gammaval} as we found in the spherical case.) Hence it appears to be the extrinsic curvature of the boundary that is responsible for the existence of these modes. 

The background geometry is given by
\begin{equation}
{\rm d}s^2 = -{\rm d}t^2 + {\rm d}r^2 + r^2 {\rm d}\varphi^2 + {\rm d}z^2\,,
\end{equation}
with $t, z \in \mathbb{R}$, $\varphi \sim \varphi + 2\pi$, and $r \geq 0$. Since $\partial/\partial t$, $\partial/\partial \varphi$, and $\partial/\partial z$ are commuting Killing vector fields, any perturbation can be expanded in terms of the combined harmonic $e^{i\omega t + i k z + i m \varphi}$. The most general metric perturbation then takes the form
\begin{multline}
h^{\rm phys}_{ab}{\rm d}x^a{\rm d}x^b\equiv e^{i\omega t+i k z+i m \varphi}\big[\hat{h}_{tt}(r)\,{\rm d}t^2+2\hat{h}_{tr}(r)\,{\rm d}t\,{\rm d}r+2\hat{h}_{tz}(r)\,{\rm d}t\,{\rm d}z+2\hat{h}_{t\varphi}(r)\,{\rm d}t\,{\rm d}\varphi
\\
\hat{h}_{rr}(r)\,{\rm d}r^2+2\hat{h}_{rz}(r)\,{\rm d}r\,{\rm d}z+2\hat{h}_{r\varphi}(r)\,{\rm d}r\,{\rm d}\varphi+\hat{h}_{zz}(r)\,{\rm d}z^2+2 \hat{h}_{z\varphi}(r)\,{\rm d}z\,{\rm d}\varphi+\hat{h}_{\varphi\varphi}(r)\,{\rm d}\varphi^2\big]\,.
\end{multline}

We choose a perturbation with $h^{\rm phys}_{ta}=0$. Imposing the Einstein equations then determines the remaining variables, and after a few algebraic manipulations, we obtain
\begin{equation}
\begin{alignedat}{1}
\hat{h}_{rr}(r) & =-\frac{k^4 r^2+k^2 \left(m^2-r^2 \omega ^2-1\right)+\left(m^2-1\right) \omega ^2}{r^2 \left(k^2-\omega ^2\right)^2}q_1(r) -\frac{q_2(r)}{r^2}+\frac{q_2'(r)}{r}\,,
\\
\hat{h}_{r\varphi}(r)& = -\frac{i \left(m^2-1\right) \left(k^2+\omega ^2\right) q_1'(r)}{m \left(k^2-\omega ^2\right)^2}-\frac{i r \left(k^2+\omega ^2\right) q_1(r)}{2 m \left(k^2-\omega ^2\right)}+q_2(r) \left[\frac{i r (k^2-\omega^2
   )}{2 m}+\frac{i m}{r}\right]-\frac{i q_2'(r)}{m}\,,
   \\
   \hat{h}_{rz}(r) & =-\frac{i k q_1'(r)}{k^2-\omega ^2}-\frac{i \left(k^2+\omega ^2\right) q_1(r)}{2 k r \left(k^2-\omega ^2\right)}+\frac{i (k^2-\omega^2)q_2(r)}{2 k r}\,,
   \\
   \hat{h}_{\varphi z}(r)&=\frac{r \left(\omega ^2-k^2\right) q_2'(r)}{2 k m}+\frac{k m q_1(r)}{k^2-\omega ^2}+\frac{\left(k^2 r+r \omega ^2\right) q_1'(r)}{2 k^3 m-2 k m \omega ^2}\,,
   \\
   \hat{h}_{z z}(r)&=q_1(r)\,,
   \\
   \hat{h}_{\varphi \varphi}(r)&=\frac{k^2 \left(m^2+r^2 \omega ^2-1\right)+\omega ^2 \left(m^2-r^2 \omega ^2-1\right)}{\left(k^2-\omega ^2\right)^2}q_1(r) -r q_2'(r)+q_2(r)\,,
\end{alignedat}
\end{equation}
where the functions $q_i(r)$, with $i = 1, 2$, satisfy the following two simple decoupled ordinary differential equations:
\begin{equation}
\frac{\left( r q_i' \right)'}{r} - \left(k^2 - \omega^2 + \frac{m^2}{r^2} \right) q_i = 0\,,
\end{equation}
and $'$ denotes differentiation with respect to $r$. Near the origin, we require that $q_i \sim r^m$, which ensures that the metric perturbation ansatz remains regular at $r = 0$ for $m \in \mathbb{Z}$.

It remains to discuss the boundary conditions at the cavity wall. As in the previous sections, we take the cavity to have unit radius.
Before imposing the boundary conditions, we gauge-transform our metric perturbation using the following gauge parameter:
\begin{equation}
\xi_a {\rm d}x^a = e^{i\omega t + i k z + i m \varphi} \left[ \delta T(r) {\rm d}t + \delta R(r) {\rm d}r + \delta \Phi(r) {\rm d}\varphi + \delta Z(r) {\rm d}z \right]\,.
\end{equation}
The behaviour of the functions $\delta T(r)$, $\delta R(r)$, $\delta \Phi(r)$, and $\delta Z(r)$ for $r < R$ is arbitrary, provided they generate a smooth gauge transformation. A sufficient condition for this is that
\begin{equation}
\delta T(r) \sim r^m\,, \quad \delta Z(r) \sim r^m\,, \quad \delta R(r) \sim r^{m-1}\quad \text{and} \quad \delta \Phi(r) \sim r^m\,,
\end{equation}
near $r=0$.

Imposing our boundary conditions, together with the requirement that the gauge-transformed perturbation satisfies $h_{tr}(1) = h_{t\varphi}(1) = h_{tz}(1) = 0$ (ensuring vanishing corner angle), determines the values of $\delta T(1)$, $\delta R(1)$, $\delta \Phi(1)$, $\delta Z(1)$, and $\delta T'(1)$, as well as the necessary boundary conditions for $q_1(r)$ and $q_2(r)$ at the cavity wall, which are given by
\begin{equation}
\begin{alignedat}{1}
&q_1'(1) - \frac{m^2 \left( 4 k^2 - 4 \omega^2 - 2 \right) + \left( k^2 - \omega^2 \right) \left( 2 k^2 + 6 p - 2 \omega^2 - 1 \right) + 2 m^4}{k^2 - \omega^2} q_1(1) = 0\,,
\\
&q_2'(1) - \frac{\left( k^2 + \omega^2 \right) \left[ m^2 \left( 4 k^2 - 4 \omega^2 - 2 \right) + \left( k^2 - \omega^2 \right) \left( 2 k^2 + 6 p - 2 \omega^2 - 1 \right) + 2 m^4 \right]}{\left( k^2 - \omega^2 \right)^3} q_1(1) = 0\,.
\end{alignedat}
\label{eq:bcscyl}
\end{equation}
The equations for the $q_i(r)$ can be readily solved in terms of Bessel functions. For smooth solutions, we require $q_i(r) \sim r^m$, which gives 
\begin{equation}
q_i(r) = A_i \, J_m(r \, \Omega) \quad \text{with} \quad \Omega \equiv \sqrt{\omega^2 - k^2}\,,
\end{equation}
where $A_i$ is a constant.

Substituting the solutions into our boundary conditions (\ref{eq:bcscyl}), yields the following two sets of possible non-trivial solutions for $\Omega$:
\begin{subequations}
\begin{equation}
J^\prime_m(\Omega) = 0\,,
\end{equation}
and 
\begin{equation}
\left[ 2 \Omega^4 - \Omega^2 \left( 4 m^2 + m + 6 p - 1 \right) + 2 m^2 \left( m^2 - 1 \right) \right] J_m(\Omega) + \Omega^3 J_{m-1}(\Omega) = 0\,.
\end{equation}
\end{subequations}
Solutions of the first equation are not of great interest to us, since one can show that they are real. However, this is not the case for the second equation.
For fixed $k$, taking $m$ to be large and using the asymptotic expansion given in Appendix \ref{app:Bessel}, we find that this equation admits solutions of the form
\be\label{eq:omegaasympcyl}
\omega = m - \Gamma \left( \frac{m}{2} \right)^{1/3} -\frac{12+60 p-53 \Gamma ^3-48 \Gamma ^6}{20 \Gamma  \left(7+16 \Gamma ^3\right)} \left( \frac{2}{m} \right)^{1/3}+\mathcal{O}\left(m^{-1} \right)\,,
\ee
with exactly the same two possible complex values for $\Gamma$ as for the case of a spherical boundary, as given in Eq.~(\ref{eq:Gammaval}). The dependence on $k$ is hidden in the $\mathcal{O}\left(m^{-1} \right)$ term. Hence we again have solutions which grow exponentially in time. Using these solutions we can repeat the argument for ill-posedness that we made above for the case of a spherical boundary. The steps of the argument are entirely analogous to those above so we will not present them here.

\section{Summary}
\label{sec:summary}

We have considered the well-posedness of linearized gravitational dynamics for a cavity with static spherical boundary whose interior is Minkowski spacetime, and where conformal, or generalized conformal, boundary conditions are imposed on this boundary.
Our starting point is the unstable perturbations about the empty cavity which were first discussed in~\cite{Anninos:2023epi} for the conformal case, and then in~\cite{Liu:2024ymn} for the generalized case. The key feature of these modes is that they are smooth, harmonic in time, with an imaginary unstable part of their frequency that scales as $\sim \ell^{1/3}$ as the angular dependence of the spherical harmonic increases, $\ell \to \infty$, and all have the same corner angle as the unperturbed cavity. Using these we have constructed a sequence of smooth solutions, with zero corner angle, with initial data that converges to zero as $\ell \to \infty$, and yet for $t>0$ the gauge invariant Riemann tensor does not converge to zero (or indeed to any smooth limit). This implies that solutions do not continuously depend on initial data, and hence the linear gravitational IBVP with conformal (or generalized conformal) boundary conditions is not well-posed. It is worth emphasizing that we have considered the gauge invariant Riemann tensor to show the solution does not converge in the large $\ell$ limit. This argument does not depend on the gauge chosen, but must hold for any gauge. We have seen that the modes localize to the ``near boundary'' region of width $\sim \ell^{-2/3}$ for large $\ell$ which indicates this lack of well-posedness is clearly associated to the presence of the (generalized) conformal boundary conditions.

Clearly our result is in tension with the recent work of
An and Anderson in~\cite{An:2025rlw}. Since we consider the linear problem, our sequence comprises solutions that are all smooth, and they all have trivial corner angle, then their well-posedness result should apply in our setting. It is obviously of great interest to understand how to resolve this tension. After sending the authors of \cite{An:2025rlw} a draft of the present paper, they informed us (M. Anderson, private communication) that their work does not establish continuous dependence of the solution on the initial-boundary-corner data, and is therefore not in disagreement with our results. This means that the work of \cite{An:2025rlw} does not prove well-posedness. Our work proves that the problem is {\it not} well-posed for linearization about a spherical (or cylindrical) boundary.

For the spherical cavity the instability growth rate becomes slower as we take the radius of the cavity $R$ to be large. Holding fixed $R/\ell$, the length scale of the perturbation on the boundary, then the growth rate scales with the cavity radius as $R^{-2/3}$.
It would be very interesting to better understand what type curvature drives this instability. Is it the fact that our spherical boundary is intrinsically curved? Or that it has non-zero mean curvature? Or some non-zero principal curvature? 
We have partial answers to these questions. We have demonstrated a similar sequence of unstable modes occurs in the interior of a timelike cylindrical boundary for (generalized) conformal boundary conditions. Since this boundary has vanishing intrinsic curvature, this suggests that it is the extrinsic curvature of the boundary that gives rise to the instability, rather than its intrinsic curvature. Furthermore we have shown a similar sequence exists for the exterior to the spherical boundary. Since the extrinsic curvature of this is opposite in sign to that for the interior, it shows that the sign of the mean curvature is not what determines stability.

Another question is to understand well-posedness of the IBVP for gravity with a cosmological constant. For a positive cosmological constant~\cite{Anninos:2024wpy} found that as the cavity wall was taken near the de Sitter horizon the time scale of the instability stopped growing with $\ell$. For negative cosmological constant one expects that the instability stops in the limit that the cavity boundary is taken to the AdS boundary, since well-posedness was proven in this context by the seminal work of Friedrich \cite{Friedrich:1995vb}, and this was indeed seen in~\cite{Anninos:2024xhc}. 
It would be interesting in future work to better understand when instabilities occur for conformal boundary conditions, and their generalization, that signal failure of well-posedness. In particular what role does the geometry of the cavity interior play, and what is the role of the geometry of the boundary?

Finally, our result implies that it is likely that the full non-linear IBVP with generalized conformal boundary conditions is not well-posed. It would be interesting to understand whether the argument presented here could be extended to analyse the full non-linear gravitational system.
Given that An and Anderson have previously argued in~\cite{An:2021fcq} that taking boundary conditions of fixing the induced metric or alternatively fixing the extrinsic curvature are not well-posed (except for the special vanishing extrinsic curvature case~\cite{Fournodavlos:2020wde}) then assuming our result extends to the full gravitational theory, this apparently leaves no natural geometric boundary conditions that yield a well-posed initial boundary value problems. Very recently they have argued that an additional open geometric condition on the extrinsic curvature may rescue well-posedness for fixed induced metric~\cite{An:2025gvr}. Perhaps the same is possible for generalized conformal boundary conditions.
Given that timelike boundaries have played a prominent role in many physical contexts such as York's regularization of the gravitational path integral~\cite{York:1986it}, holographic renormalization~\cite{Henningson:1998gx,deHaro:2000vlm,Skenderis:2002wp}, the fluid-gravity paradigm~\cite{Bredberg:2011xw,Anninos:2011zn}, and many more recent discussions in AdS-CFT
such as $T\bar{T}$ deformations~\cite{McGough:2016lol,Kraus:2018xrn,Hartman:2018tkw,Gorbenko:2018oov},
this lack of well-posed Lorentzian evolution is surprising and clearly requires further study. 

\section*{Acknowledgments} 

We thank Michael Anderson for comments on a draft and Claude Warnick for a helpful discussion. We also thank Dionysios Anninos for interesting discourse. HSR and JES are supported by STFC grant No. ST/X000664/1 and TW by ST/T000791/1. JES is partially supported by Hughes Hall College. XL is supported in part by NSF grant PHY-2408110 and by funds from the University of California.

\appendix

\section{Further details of the metric perturbation}
\label{app:smoothness}

In this appendix we present some details of the metric perturbation.

\subsection{Perturbation components}

The scalar perturbation $h^{\rm phys}_{ab}$ in~\eqref{eq:scalar} has the following components;
\be
\begin{split}
&
h^{\rm phys}_{tt}  = - e^{i \omega t}  f(\theta)  \omega^2 \delta h_T(r) 
\\
&
h^{\rm phys}_{tr}  = i \omega e^{i \omega t}  f(\theta)   \left( \delta h_T'(r) - \frac{1}{r} \delta h_T(r) - \delta h_r(r) \right)
\\
&
h^{\rm phys}_{rr}  = e^{i \omega t}  f(\theta) \left(  \delta b(r) - 2 \delta h_r'(r) + \delta h_T''(r) - \frac{2}{r} \delta h_T'(r) + \frac{2}{r^2} \delta h_T(r) \right)
\\
&
h^{\rm phys}_{IJ}  = e^{i \omega t}  f(\theta)  \left( \delta h_L(r) - 2 r \delta h_r(r)  + r \delta h_T'(r) + \frac{\ell^2 + \ell - 4}{2} \delta h_T(r) \right) \Omega_{IJ} \; .
\end{split}
\ee
Further, the gauge part $h^{\rm gauge}_{ab}$ in~\eqref{eq:gauge} has components;
\begin{equation}
\begin{split}
&h^{\rm gauge}_{tt}=- i \omega \delta T(r)\,f(\theta)\,e^{i \omega t}
\\
&h^{\rm gauge}_{tr}=\frac{1}{2} \left[ i \omega r \delta R(r) - \delta T'(r) \right] f(\theta)\,e^{i \omega t}
\\
&h^{\rm gauge}_{tI}=\frac{1}{2} \left[ i \omega r^2 \delta Q(r) - \delta T(r) \right]  D_I  f(\theta)\,e^{i \omega t}
\\
&h^{\rm gauge}_{rr}=[r\delta R'(r)+\delta R(r) ] f(\theta)\,e^{i \omega t}
\\
&h^{\rm gauge}_{rI}=\frac{1}{2} \left[r \delta R(r) + r^2 \delta Q'(r) \right] D_I  f(\theta)\,e^{i \omega t}
\\
&h^{\rm gauge}_{IJ}=\left[r^2 \delta R(r)- \frac{1}{2} \ell ( \ell + 1 ) r^2 \delta Q(r)\right]  f(\theta)\, \Omega_{IJ}\,e^{i \omega t} +  r^2 \delta Q(r)\,S_{IJ}\,e^{i \omega t} \; .
\end{split}
\end{equation}

\subsection{Smoothness of the perturbation at the origin}

Consider our scalar perturbation, $h^{\rm phys}_{ab}$, given by~\eqref{eq:scalar} for our modes with~\eqref{eq:sol} and solving the equation~\eqref{eq:beqn} for $\delta b(r)$. We will explicitly demonstrate this is a smooth tensor at the origin once transformed to Cartesian coordinates where the  components  are then smooth functions.
We transform to Cartesian coordinates $x^{a'} = (t, x, y, z)$ in the usual manner,
\be
(x,y,z) = (r \sin{\theta} \cos{\phi}, r \sin{\theta} \sin{\phi}, r \cos{\theta} ) 
\ee
and then in these new coordinates,
\be
h^{\rm phys}_{a'b'} &=& \frac{\omega^2}{\ell ( \ell + 1) ( \ell^2 + \ell - 2 ) } M_{a'b'} \;,
\ee
where writing $x^{i'} = (x,y,z)$ then,
\be
\label{eq:metcart}
M_{t't'} = A(t,r,\theta) + r^2 B(t,r,\theta) \; , \quad M_{i'j'} = \delta_{i'j'} A(t,r,\theta) + x^{i'} x^{j'} B(t,r,\theta)
\; \quad M_{t'i'} = x^{i'} C(t,r,\theta) \nl
\ee
with,
\be
A(t,r,\theta) &=& - \left( (6 + \ell + \ell^2) \Phi(t,r,\theta) + 2 \Pi(t,r,\theta) \right)  \nl
B(t,r,\theta) &=& 2 \omega^2 \Phi(t,r,\theta) \nl
C(t,r,\theta) &=& - 2 i \omega \left( 4 \Phi(t,r,\theta) + \Pi(t,r,\theta) \right)\;,
\ee
where we have defined,
\be
\label{eq:phipi}
\Phi(t,r,\theta) = r^2 \delta b(r) f(\theta) e^{i \omega t} \; ,\qquad \Pi(t,r,\theta) = r^3 \delta b'(r) f(\theta) e^{i \omega t} \; .
\ee
In computing the above we have used the equation for $\delta b$ in~\eqref{eq:beqn}, as well as that for the spherical harmonic  $f(\theta) = \sqrt{\frac{4\pi}{2\ell+1}} P_\ell(\cos \theta)$ which obeys
$f''(\theta) + \cot\theta f'(\theta) + \ell( \ell+1) f(\theta) = 0$.
Since $f(\theta)$ is a Legendre polynomial in $\cos\theta$ of order $\ell$, so (for integer $\ell$),
\be
r^\ell f(\theta) = \, r^\ell P_\ell(\cos\theta) = \left\{ 
\begin{array}{cc}
a_\ell z^\ell  + a_{\ell-2} r^2 z^{\ell-2} + \ldots + a_0 (r^2)^{\frac{\ell}{2}} & \quad \ell \; \mathrm{even}\\
a_\ell z^\ell + a_{\ell-2} r^2 z^{\ell-2} + \ldots + a_1 (r^2)^{\frac{\ell - 1}{2}} z  & \quad \ell \; \mathrm{odd}
\end{array}
\right. \; .
\ee
As a consequence, any function $F(r)$ that is a smooth function of $r^2$ multiplied by this, so $F(r) r^\ell f(\theta)$, is a smooth functions of the Cartesian coordinates.

Now for integer $\ell$ then from~\eqref{eq:soln} we see that $\delta b(r) = r^{\ell-2} F(r)$ for $F(r)$ that is smooth in $r^2$. Hence we see that the functions $\Phi(t,r,\theta)$ and $\Pi(t,r,\theta)$ defined in~\eqref{eq:phipi} above are  smooth in Cartesian coordinates. As a consequence, we see that the components $M_{a'b'}$ given above are likewise smooth, and hence the components $h^{\rm phys}_{a'b'}$ are too. Thus we see that the physical part of the perturbation is smooth at the origin.

We may also consider the vector field $\xi^a$ defined in~\eqref{eq:xi}. In Cartesian coordinates this has components, 
\be
\label{eq:xicart}
\xi^{t'} &=& \frac{1}{2} e^{i \omega t} \delta T(r) f(\theta) \nl
\xi^{i'} &=& \frac{1}{2} x^{i'}  e^{i \omega t} \delta R(r) f(\theta) 
+ \frac{1}{2} e^{i \omega t} ( x^{i'} z - r^2 \delta_{i' z} )\delta Q(r) \frac{f'(\theta)}{r \sin(\theta)} \; .
\ee
Noting that,
\be
r^\ell \frac{f'(\theta)}{r \sin{\theta}}  = \left\{ 
\begin{array}{cc}
a_{\ell-1} z^{\ell-1} + a_{\ell-3} r^2 z^{\ell-3} + \ldots + a_1 (r^2)^{\frac{\ell - 2}{2}} z & \quad \ell \; \mathrm{even}\\
a_{\ell-1} z^{\ell-1}  + a_{\ell-3} r^2 z^{\ell-3} + \ldots + a_0 (r^2)^{\frac{\ell-1}{2}} & \quad \ell \; \mathrm{odd}
\end{array}
\right.
\ee
then to ensure that these components $\xi^{a'}$ are smooth functions of the Cartesian coordinates a sufficient condition is to require that $\delta T(r) = r^\ell F_T(r)$, $\delta R(r) = r^{\ell} F_R(r)$ and $\delta Q(r) = r^\ell F_Q(r)$ where $F_{T,R,Q}(r)$ are smooth functions of $r^2$, hence the scalings in equation~\eqref{eq:smoothgauge}. We note that this condition is not necessary -- one can allow scalings with a lower power than $\sim r^\ell$ if $\delta R$ and $\delta Q$ are tuned to give a suitable cancellation between the two terms involving them in the components $\xi^{i'}$ above.

\section{Bessel function approximation for large order}
\label{app:Bessel}

Our starting point is the asymptotic approximation at large order of Olver~\cite{Olver1954,NIST:DLMF} which states that for real $\nu$ and $a \in \mathbb{C}$ then $J_{\nu}\left(\nu + a \nu^{1/3} \right)$ has an asymptotic expansion,
\be
J_{\nu}\left(\nu + a \nu^{1/3} \right) & \sim &  \left( \frac{2}{\nu} \right)^{1/3} \mathrm{Ai}(- 2^{1/3} a) \sum_{n=0}^\infty \frac{P_n(\alpha)}{\nu^{2n/3}} + \frac{2^{2/3}}{\nu} \mathrm{Ai}'(- 2^{1/3}  a) \sum_{n=0}^\infty \frac{Q_n(\alpha)}{\nu^{2n/3}}  
\ee
as $\nu \to \infty$ that is uniform with respect to $a$. Here $\mathrm{Ai}$ is the Airy function of the first kind, $P_n(\alpha)$ and $Q_n(\alpha)$ are polynomials in $\alpha$, and
the leading terms are given by,
\begin{multline}\label{eq:Pn}
P_0 = 1 \; , \quad P_1 = -\frac{a}{5} \; , \quad P_2 = - \frac{9 a^5}{100} + \frac{3 a^2}{35} \; , \quad P_3 =  \frac{957 a^6}{7000} - \frac{173 a^3}{3150} - \frac{1}{225} \; ,
\\
P_4 = \frac{27 a^{10}}{20000} - \frac{23573 a^7}{147000} + \frac{5903 a^4}{138600} + \frac{947 a}{346500}\,,
\end{multline}
and,
\begin{multline}\label{eq:Qn}
Q_0 = \frac{3 a^2}{10} \; , \quad Q_1 =  - \frac{17 a^3}{70} + \frac{1}{70} \; , \quad Q_2 = - \frac{9 a^7}{1000} + \frac{611 a^4}{3150} - \frac{37 a}{3150} \; ,
\\
Q_3 = \frac{549 a^8}{28000} - \frac{110767 a^5}{693000} + \frac{79 a^2}{12375}\,.
\end{multline}
For real $q$, writing $\nu = q + \ell$ and defining $\alpha \in \mathbb{C}$ through $\nu + a \nu^{1/3} = \ell - \alpha \left( \frac{\ell}{2} \right)^{1/3}$ then we have,
\be
- 2^{1/3} a =  \alpha  + 2^{1/3} q \frac{1}{\ell^{1/3}} - \frac{q \alpha}{3} \frac{1}{\ell} - \frac{2^{1/3} q^2}{3} \frac{1}{\ell^{4/3}}
+ \frac{2 q^2 \alpha}{9} \frac{1}{\ell^2}+ \frac{2^{4/3} q^3}{9} \frac{1}{\ell^{7/3}} - \frac{14 q^3 \alpha}{81} \frac{1}{\ell^3}
+ O\left( \ell^{-10/3}\right) \; . \nl
\ee
Further using that the Airy function is entire, then $\mathrm{Ai}(x + \delta x)$ equals its Taylor expansion so,
\be
\mathrm{Ai}(x + \delta x) = \sum_{n=0}^\infty \mathrm{Ai}^{(n)}(x) \cdot (\delta x)^n/n! = \sum_{n=0}^\infty \left[ \mathrm{Ai}(x)  p_n(x) +  \mathrm{Ai}'(x)  q_n(x) \right] (\delta x)^n
\ee
for polynomials $p_n(x)$, $q_n(x)$ using the Airy equation $\mathrm{Ai}''(x)=x\,\mathrm{Ai}(x)$, with a similar expression also holding for $\mathrm{Ai}'(x + \delta x)$. 
Hence for $\alpha \in \mathbb{C}$ we may write the asymptotic expansion,
\be
\label{eq:BesselApprox}
J_{q+\ell}\left(\ell - \alpha \left( \frac{\ell}{2} \right)^{1/3} \right) & \sim &  \mathrm{Ai}(\alpha) \sum_{n=1}^\infty \tilde{P}_n(q,\alpha) \left( \frac{2}{\ell} \right)^{n/3} + \mathrm{Ai}'(\alpha) \sum_{n=2}^\infty \tilde{Q}_n(q,\alpha) \left( \frac{2}{\ell} \right)^{n/3} 
\ee
for $\ell \to \infty$ which is uniform with respect to $\alpha$, where $\tilde{P}_n(\alpha,q)$ and $\tilde{Q}_n(\alpha,q)$ are polynomials in $\alpha$ and $q$ of order $n$ and $n-1$, respectively. Explicitly, the leading terms are given by
\be
\tilde{P}_1(\alpha,q) = 1 \, , \quad \tilde{P}_2(\alpha,q) = 0 \, ,
\quad \tilde{P}_3(\alpha,q) = \frac{1+5q^2}{10}\alpha \, , \quad \tilde{P}_4(\alpha,q) =  \frac{9 \alpha^3 + 10 q^2 - 4}{60}q\,,
\ee
and
\be
\tilde{Q}_2(\alpha,q) = q \; , \quad \tilde{Q}_3(\alpha,q) = \frac{3}{20} \alpha^2 \; , \quad \tilde{Q}_4(\alpha,q) =  \frac{ 7 + 5 q^2}{30}\,q\,\alpha\,.
\ee
There is a similar expansion for the Hankel function of the second kind, namely,
\begin{equation}
    H_{\nu}\left(\nu + a \nu^{1/3} \right)  \sim   2\left( \frac{2}{\nu} \right)^{\frac{1}{3}}e^{\frac{i\pi}{3}} \mathrm{Ai}(e^{\frac{i\pi}{3}} 2^{\frac{1}{3}} a) \sum_{n=0}^\infty \frac{P_n(\alpha)}{\nu^{2n/3}} + \frac{2^{\frac{5}{3}}}{\nu} e^{-\frac{i\pi}{3}}\mathrm{Ai}'(e^{\frac{i\pi}{3}} 2^{\frac{1}{3}}  a) \sum_{n=0}^\infty \frac{Q_n(\alpha)}{\nu^{2n/3}}  \,,
\end{equation}
where $P_n$ and $Q_n$ are given in~\eqref{eq:Pn} and~\eqref{eq:Qn}.

\section{Proof of existence of unstable perturbations for large $\ell$}
\label{app:omega}

We begin by writing the equation~\eqref{eq:omega} determining $\omega$ as,
\be
F(\omega) &=& P_\ell(\omega) J_{\frac{3}{2}+\ell}(\omega) -  Q_\ell(\omega)  J_{\frac{1}{2}+\ell}(\omega)  
\ee
where  $P_\ell(\omega), Q_\ell(\omega)$ are polynomials in $\omega$,
\be
P_\ell(\omega) &=& \omega \left[ 3 \ell ( \ell+1) + 24 p - 2(2 + \omega^2) \right]  \nl
 Q_\ell(\omega) &=&  (\ell+1)(\ell+2)(2 \ell^2 + \ell - 2 + 12 p) - 2 ( 2 \ell^2 + 3 \ell - 1 + 12 p) \omega^2 + 2 \omega^4 \; .
\ee
Now let us write the equation~\eqref{eq:gammaeqn} that $\Gamma$ satisfies as,
\be
H(a)  =  \mathrm{Ai}'(a)   -  4  a^2 \mathrm{Ai}(a)
\ee
so that $H(\Gamma) = 0$ with $\Gamma$ as in~\eqref{eq:Gammaval}. 
\footnote{
We may numerically verify the existence of this root to high accuracy. One way is to directly solve for the root. Another is to numerically compute $\frac{1}{2 \pi i} \oint \frac{da}{H(a)}$ for different curves enclosing only the root $\Gamma$ and show that it consistently generates a residue -- we find the residue to be $\mathrm{Res}_\Gamma \simeq -0.223334 + 0.735103 i$. 
}
Given that this root $a = \Gamma$ exists, we will now prove  that for sufficiently large $\ell$ there always exist solutions with $\omega \simeq \ell - \Gamma \left( \frac{\ell}{2} \right)^{1/3}$.

We begin by writing $\omega = \ell - a \left( \frac{\ell}{2} \right)^{1/3}$ so that $P_\ell(\omega)$ and $Q_\ell(\omega)$ are polynomials in $a$. Then,
\be
 P_\ell(\omega) = \ell^3 \left[ 1  + \frac{\delta P_\ell(a)}{\ell^{2/3}}  \right]  \; , \qquad Q_\ell(\omega) =  \ell^3  \left[ 1 + 4  a^2 \left(  \frac{2}{\ell} \right)^{1/3} + \frac{\delta Q_\ell(a)}{\ell^{2/3}}  \right]\;,
 \ee
 where $\delta P_\ell(a), \delta Q_\ell(a)$ are polynomials in $a$ and $\delta P_\ell(a), \delta Q_\ell(a) \sim \mathcal{O}(1)$ as $\ell \to \infty$.
 Now from our asymptotic expansions we have;
 \be
 \label{eq:jexp}
 \begin{split}
     J_{\frac{3}{2}+\ell}(\omega)  &= \frac{2^{1/3} \mathrm{Ai}(a) }{\ell^{1/3}} +  \frac{3 \mathrm{Ai}'(a)  }{2^{1/3} \ell^{2/3}}  +  \frac{\delta {J}_\ell(a)}{\ell} 
\; , \\
  J_{\frac{1}{2}+\ell}(\omega)  &= \frac{2^{1/3} \mathrm{Ai}(a) }{\ell^{1/3}} +  \frac{\mathrm{Ai}'(a)  }{2^{1/3} \ell^{2/3}}  +  \frac{\delta \tilde{J}_\ell(a)}{\ell} \;,
 \end{split}
\ee
so that $\delta J_{\ell}(a)$ and $\delta \tilde{J}_{\ell}(a)$ are uniformly bounded as $\delta J_{\ell}(a), \delta \tilde{J}_{\ell}(a) \sim \mathcal{O}(1)$ uniformly in $a$ as $\ell \to \infty$. Now an important point is that since the Bessel function for fixed order is an entire function of its argument, and likewise the Airy function and its derivative are entire functions, therefore the remainders $\delta J_{\ell}(a)$ and $\delta \tilde{J}_{\ell}(a)$ are also entire functions of $a$. 
Then defining $\tilde{F}(a) =  \ell^{-7/3} F(\omega)$ then,
\be
\tilde{F}(a) &=& 2^{2/3} H(a) + R(a)\;,
\ee
where the `remainder' is,
\be
&& R(a) =   \frac{\delta J_\ell(a) -  \delta \tilde{J}_\ell(a)  + 2^{1/3} [ \delta P_\ell(a) - \delta Q_\ell(a) ] \mathrm{Ai}(a)    -  4  a^2   \mathrm{Ai}'(a) }{\ell^{1/3}} \nl
&&   \qquad\qquad
+ \frac{  2^{-1/3} [3 \delta P_\ell(a) - \delta Q_\ell(a) ] \mathrm{Ai}'(a)   -  4  a^2 2^{1/3}   \delta \tilde{J}_\ell(a)  }{\ell^{2/3}} \nl
&&\qquad \qquad
  + \frac{ \delta P_\ell(a) \delta J_\ell(a) - \delta Q_\ell(a)   \delta \tilde{J}_\ell(a) }{\ell} \;.
 \ee
Let $\epsilon > 0$. We define a curve $\mathcal{C}$ in the complex $a$-plane as $a = \Gamma + \epsilon \, e^{i \theta}$ for angle $\theta$. We restrict ourselves to sufficiently small $\epsilon$  so that the curve encircles the single root of $H(a)$ at $a = \Gamma$ and does not encircle or meet any other roots. Let $c = \min_{\mathcal{C}} | H(a) |$, and we note $c > 0$.
Denote the region $\mathcal{R}$ as all $a$ such that $| a | \le | \Gamma | + \epsilon$, and note that the curve $\mathcal{C}$ lies within $\mathcal{R}$.

From the discussion above, there exists constants $k$ and $\ell_0$ such that for all $a \in \mathcal{R}$ and $\ell > \ell_0$ we have, 
\be
|a|, | \mathrm{Ai}(a) |, | \mathrm{Ai}'(a) |,  | \delta P_\ell(a) |, | \delta Q_\ell(a) |, | \delta J_\ell(a) |, |\delta \tilde{J}_\ell(a) | < k\;.
\ee
Hence from the form of the remainder, $R(a)$, there exists $\ell_1 \ge \ell_0$ such that for $\ell > \ell_1$ then $| R(a) | < c$ for all $a \in \mathcal{C}$.

Both $H(a)$ and $R(a)$ are entire functions of $a$. For $\ell > \ell_1$ we have that $| R(a) | < | 2^{2/3} H(a) |$ on $\mathcal{C}$. Then Rouche's theorem implies that the number of roots of $H(a)$ within $\mathcal{C}$ (which by construction is just the one at $a = \Gamma$) is the same as that of $\tilde{F}(a) = 2^{2/3} H(a) + R(a)$. Hence we have proved the existence of a root of $\tilde{F}(a)$ for sufficiently large $\ell$ which is a maximum distance $\epsilon$ from $\Gamma$. Taking small $\epsilon$ this implies corresponding roots of $F(\omega)$ for all $\ell > \ell_1$ that have $\omega = \ell - a \left( \frac{\ell}{2} \right)^{1/3} \simeq \ell - \Gamma \left( \frac{\ell}{2} \right)^{1/3}$.

\section{Proofs of lemma in Section~\ref{sec:bounds}}
\label{app:proofs}

Here we give a proof of the bound stated in the lemma of Section~\ref{sec:bounds}.
Given our function $u(r)$ defined in~\eqref{eq:fnu} we have two propositions which we prove shortly;

\begin{prop}
Given $R_0 > 0$ there exists $\ell_0$ such that for $\ell_0 < \ell$ the magnitude of $u(r)$ in the domain $r \in [0,1]$ is bounded by its magnitude in the near boundary region $0 \le \delta r \le R_0$.
\end{prop}

\begin{prop}
Given $R_1 > 0$ there exists $\ell_1$ and $C$ such that in the near boundary region $0 \le \delta r \le R_1$ for $\ell_1 < \ell$ then $\left| u(r) \right| < C$.
\end{prop}

\noindent
As a corollary of these, taking $R_1 = R_0$ we have that for $\max(\ell_0,\ell_1) < \ell$ then $| u(r) | < C$  over the whole domain $r \in [0, 1]$ which proves the lemma.
In the following two sections we prove these two propositions.

\subsection{Proposition 1: Maximum principle}

As in Section~\ref{sec:bounds} we take the function $u(r)$ to be;
\be
u(r) = \frac{\ell^{1/3}}{r^{q + a}} J_{q + \ell}(r \omega)
\ee
on our domain $r \in [0,1]$ with $q$ real, and $a, \ell$ integers, and $\omega$ given as a solution to~\eqref{eq:omega} so that at large $\ell$ it has the behaviour~\eqref{eq:omegaasymp}. We further restrict to $\ell > a$ and then $u(r)$ is smooth over the whole domain, and $u(0) = 0$.
Then this obeys the equation, 
\be
\partial^2_r u(r) + \frac{1 + 2 a + 2 q}{r} \partial_r u(r) + \left( \frac{r^2 \omega^2 - ( \ell - a) (\ell + a + 2 q ) }{r^2}\right) u(r) = 0 \; .
\ee
Now consider the magnitude squared of this,
\be
F(r) = \left| u(r) \right|^2
\ee
which is positive so $F(r) > 0$.
Then this obeys,
\be
\label{eq:F}
F''(r) + \frac{1 + 2 a + 2 q}{r} F'(r) + 2 \ell^2 \Omega(r) F(r) = 2 \left| u'(r) \right|^2 \ge 0 
\ee
where the real function $\Omega(r)$ is given as,
\be
\Omega(r) = \frac{r^2 (\mathrm{Re}(\omega)^2 - \mathrm{Im}(\omega)^2) - ( \ell - a) (\ell + a + 2 q ) }{\ell^2 r^2}
\ee
so that at large $\ell$ it behaves as,
\be
\Omega(r) = \left( 1 - \frac{1}{r^2} \right) - \Gamma_r \left( \frac{2}{\ell} \right)^{2/3} + \mathcal{O}\left( \frac{1}{\ell} \right) \; .
\ee
Note that for $0 < r \le 1$ the first term on the righthand side is negative, but the second is positive as $\Gamma_r < 0$.
Therefore for any $R_0 > 0$ there exists $\ell_0$ such that for all $\ell > \ell_0$ and $r$ in the range,
\be
0 <  r  <  1 - \frac{R_0}{\ell^{2/3} } 
\ee
then $\Omega(r)$ is negative. Then in this range $F(r)$ can have no maximum -- at a stationary point $F'(r) = 0$ but then~\eqref{eq:F} implies $F''(r) > 0$. Now since $u(0) = 0$, then $F(0) = 0$. Hence $F(r)$ must increase from zero as $r$ increases from zero provided $r  <  1 - \frac{R_0}{\ell^{2/3}}$.

In terms of our near boundary coordinate, the condition $r  <  1 - \frac{R_0}{\ell^{2/3}}$ is simply $\delta r > R_0$. Then a corollary of the above is that
for $\ell_0 < \ell$ the magnitude of $u(r)$ in the domain $r \in [0,1]$ is bounded by its magnitude in the near boundary region $0 \le \delta r \le R_0$. \qedsymbol

\subsection{Proposition 2: Near boundary behaviour}

Using~\eqref{eq:alphar} then given some real $R_1 > 0$  there exists $A > 0$ and $\ell_1$ such that for $\ell_1 < \ell$ then $| \alpha(\delta r) | < A$ for all  $0\le\delta r\le R_1$.

Consider the asymptotic approximation above in equation~\eqref{eq:BesselApprox} for real $q$ and $\alpha \in \mathbb{C}$.
Since the approximation is uniform with respect to $\alpha$, then for any $A > 0$, we may find a constant $C$ and $\ell_2$ such that for all $| \alpha | < A$ and $\ell_2 < \ell$ then,
\be
\left| J_{q + \ell}\left(\ell - \alpha \left( \frac{\ell}{2} \right)^{1/3} \right) \right| < \frac{C}{\ell^{1/3}} \; .
\ee
We note that this uses the fact that the Airy function $\mathrm{Ai}(\alpha)$ is an entire function, and so is bounded for $| \alpha | < A$. 
Hence we see that for $\max(\ell_1, \ell_2) < \ell$ then for all $0 \le \delta r \le R_1$ we have,
\be
\left| J_{q + \ell}\left( r \omega \right) \right| < \frac{C}{\ell^{1/3}} \; .
\ee
For the same range $0 \le \delta r \le R_1$ we have, ${1}/{r^{q + a}} = 1 + O\left( \frac{1}{\ell^{2/3}} \right)$.
Thus we may find an $\ell_3$ and constant $C'$ such that for $\ell_3 < \ell$ then,
\be
\left| u(r) \right| = \frac{\ell^{1/3}}{r^{q + a}} \left| J_{q + \ell}\left( r \omega \right) \right| < C'
\ee
for all $r$ in the near boundary range defined by $0 \le \delta r \le R_1$. \qedsymbol

\nocite{*}

\bibliography{references.bib}{}
\bibliographystyle{utphys-modified}

\end{document}